\documentclass[fleqn,usenatbib]{mnras}

\usepackage[T1]{fontenc}

\DeclareRobustCommand{\VAN}[3]{#2}
\let\VANthebibliography\thebibliography
\def\thebibliography{\DeclareRobustCommand{\VAN}[3]{##3}\VANthebibliography}

\usepackage{graphicx}	
\usepackage{amsmath}	
\usepackage{amssymb}	
\usepackage{siunitx}
\usepackage{float}
\usepackage[normalem]{ulem}
\usepackage{newtxtext,newtxmath}

\newcommand{\oiii}{[{\sc O iii}]$\lambda$5007}
\newcommand{\loiii}{$L_{[\mathrm{O} \textsc{iii}]\lambda5007}$}
\newcommand{\qsfit}{{\sc QSFit}}

\title[QSFit: Type 2 AGN]{Efficient Analysis Routines for Single and Double Peaked Type 2 AGN Spectra}

\author[M. Selwood et al.]{
M. Selwood,$^{1}$\thanks{E-mail: matthew.selwood@bristol.ac.uk}
G. Calderone,$^{2}$
S. Fotopoulou,$^{1}$
M. N. Bremer$^{1}$
\\
$^{1}$H.H. Wills Physics Laboratory, Tyndall Avenue, Bristol BS8 1TL, UK\\
$^{2}$INAF -- Osservatorio Astronomico di Trieste, Via G.B. Tiepolo 11, 34143 Trieste, Italy\\
}

\date{Accepted XXX. Received YYY; in original form ZZZ}

\pubyear{2022}

\begin{document}
\label{firstpage}
\pagerange{\pageref{firstpage}--\pageref{lastpage}}
\maketitle

\begin{abstract}
Driven by the imminent need to rapidly process and classify millions of AGN spectra drawn from next generation astronomical facilities, we present a spectral fitting routine for Type 2 AGN  spectra optimised for high volume processing, using the \textit{Quasar Spectral Fitting} library (\qsfit{}). We analyse an optically selected sample of 813 luminous Type 2 AGN spectra at $z < 0.83$ from the Sloan Digital Sky Survey (SDSS) to qualify its performance. We report a median narrow line H$\alpha$/H$\beta$ Balmer decrement of 4.5$\pm$0.8, alluding to the presence of dust in the narrow line region (NLR). We publish a specialised \qsfit{} fitting routine for high signal to noise ratio spectra and general fitting routine for double peaked Type 2 AGN spectra applied on a sub-sample of 45 spectra from our parent sample. We report a median red and blue peak velocity separation of 390$\pm$60~kms$^{-1}$. No trend is found for red or blue peaks to exhibit systematically different luminosity or ionization properties. Emission line diagnostics show that the double peaks in all sources are illuminated by an AGN-powered ionizing continuum. Finally, we examine the morphology of host galaxies of our double peaked sample. We find double peaked Type 2 AGN reside in merging systems at a comparable frequency to single peaked AGN. This suggests that the double peaked AGN phenomenon is likely to have a bi-conical outflow origin in the majority of cases. We publicly release the code used for spectral analysis and produced catalogues used in this work.
\end{abstract}




\section{Introduction}
Active galactic nuclei (AGN) are a phase of luminous accretion of matter onto a central supermassive black hole (SMBH) in a galaxy, signifying the growth of the massive compact object. A significant fraction of the accreted matter rest-mass energy is converted to electromagnetic radiation that, through a number of mechanisms, is observed from radio to gamma ray bands. Observational and cosmological simulation-based works suggest that the properties and evolution of SMBH and AGN host galaxies are closely related \citep[][and references therein]{alexander&hickox2012}. Accurate determination of AGN emission line and energetic properties, which are encoded in the spectra of the objects is therefore essential for the study of this co-evolution.  

AGN are classified optically as Type 1 (unobscured) and Type 2 (obscured) based on the presence or lack of broad ($\gtrsim 10^3$ km s$^{-1}$) emission lines in their spectra. The unification model of AGN postulates that all AGN have an ubiquitous central structure comprising of several distinct components \citep{antonucci1993, urry&padovani1995, netzer2015}. In this formalism, many of the discrepancies in spectral appearance between different AGN species arise from the orientation of a standard AGN structure with respect to the observer \citep[see][for a review]{netzer2015}. Type 2 AGN are therefore proposed to have their central engine, i.e. accretion disk and broad line region (BLR), obscured from the observer by an axisymmetric dusty molecular torus located 0.1 - \SI{10}{pc} from the central SMBH.

Complete extinction of optical emission from the BLR requires a dust obscuring screen of $A_{V}$ = 5 - 10$\ $mags \citep{schnorr-muller2016}, which corresponds to an X-ray absorbing hydrogen column density of $N_{H} > 10^{22}$\ {cm$^{-2}$} using typical dust-to-gas ratios  \citep[e.g.][]{predehl&schmitt1995}. Hence, Type 2 AGN have severely diminished optical, ultraviolet (UV) and soft X-ray emission, which originate from the central few parsecs of the AGN, compared to their Type 1 counterparts. This obscuration introduces difficulty into the identification of Type 2 AGN in these bands, particularly as host-galaxy dilution can compete with and in extreme cases overwhelm AGN emission in lower luminosity sources \citep{hickox&alexander2018}.

Type 2 AGN nevertheless exhibit strong narrow optical emission lines with velocity full-width at half-maximum (FWHM) values of a few $\sim$ 100 - 1000~kms$^{-1}$. These emission lines originate from forbidden transitions in the low density narrow-line region (NLR) of AGN. The NLR extends beyond the influence of the obscuring medium, up to a few kiloparsecs from the central SMBH. 

Despite successfully reconciling many features of Type 1 and Type 2 AGN, there is significant evidence to suggest that the standard unified orientation model does not fully explain the properties observed between the optical subtypes. The host-galaxy properties between the two classes have been found to deviate, with \citet{zou2019} finding hints that Type 2 AGN have higher host-galaxy stellar masses than Type 1 AGN. The works of \citet{bornancini&garcialambas2020} and \citet{villarroel&korn2014} proposed that Type 1 and Type 2 AGN are found in different environments with distinct neighbours, favouring an evolutionary scenario between Type 1 and Type 2 AGN. Similarly, the mid-IR luminosity function study of \citet{lacy2015} shows that Type 1 and Type 2 AGN populations follow significantly different evolutionary trends through cosmic time, with an earlier peak in space density for Type 2 AGN. It is therefore of importance to characterise the spectroscopic and physical properties of samples of Type 2 AGN in order to understand their standing in relation to their Type 1 counterparts. 

While early work concentrated on single AGN spectra or a handful of sources, the introduction of large scale spectroscopic surveys such as the Sloan Digital Sky Survey \citep[SDSS,][]{blanton2017} and in years ahead Euclid \citep{laureijs2011}, DESI \citep{desi2016}, 4MOST \citep{dejong2019} and MOONS \citep{cirasuolo2020} have shifted the field towards the analysis of large samples of thousands to millions of spectra \citep[Selwood et al., in prep;][]{merloni2019, dey2019}. The volume of data from such large scale surveys has borne a plethora of codes, software distributions and recipes to analyse and decompose the properties of optical AGN spectra. Each of these routines serve a different purpose, with some codes focussing on fitting the spectra of particular types of AGN. Others are written for the extraction of specific spectral properties (eg. SFR, velocity dispersion) and some with the aim of providing precise decompositions of a given feature or conversely for high performance on large spectroscopic samples. This software ecosystem has obvious shortcomings such as the lack of transparency and reproducibility stemming from analysis codes being kept private or being close source. Many available codes are unable to treat both Type 1 and Type 2 AGN spectral fitting. As different fitting routines use varied models and algorithms systematic differences in their measurements can appear, which are often overlooked.

\textit{Quasar Spectral Fitting} library \citep[\qsfit,][]{calderone2017}) is a software package aiming to perform high volume automatic analysis of Type 1 AGN and QSO optical spectra. Originally implemented in {\sc IDL}, it has recently been ported\footnote{\url{https://github.com/gcalderone/QSFit.jl/}} in Julia language \citep{bazanson2012} to simplify analysis customization and allowing easier sharing of the recipes and of the results (Calderone et al., in prep.). The software provides a scalable means to consistently analyse large spectroscopic samples of spectra, by adopting a flexible fitting recipe which incorporates several, physically motivated, AGN spectral components to model the data. The package aims to be as general as possible providing acceptable fits in the vast majority of cases, and being customisable enough to deal with the more exotic ones. 

The \qsfit{} recipe iteratively add the model components while performing a fit following a Levenberg–Marquardt least-squares minimization algorithm \citep[][]{markwardt2009}, until all component are simultaneously fit to the data, providing an overall coherent picture of the spectral quantities. It is agnostic to the instrument used to collect the spectra, allowing the user to define an instrumental resolution.

In this work, we developed a new spectral fitting recipe specifically targeted to Type 2 AGN. We have applied our procedure to the \citet{reyes2008} sample (Sect. \ref{sec:qsfit_t2}) of high luminosity Type 2 AGN SDSS spectra (Sect. \ref{sec:sample}) in order to demonstrate the software's utility and explore specific issues around the interpretation of such spectra. We discuss self-consistency checks, consistency with the literature and initial scientific results of our sample in Sect. \ref{sec:sample_results}. As a consequence of our analysis we also develop specialised approaches to analysing Type 2 AGN spectra with high signal to noise ratios (S/N, Sect. \ref{sec:snratio}) and double peaked emission lines (Sect. \ref{sec:dp})\footnote{All codes and recipes used for spectral analysis in this work are publicly available at \url{https://github.com/MattSelwood/Type2AGN} with instructions to reproduce our analysis.}.


Throughout this paper we assume a $\Lambda$CDM cosmology with $H_{0}$ = 70$\ $kms$^{-1}$Mpc$^{-1}$, $\Omega_{m}$ = 0.3 and $\Omega_{\Lambda}$ = 0.7.

\section{Type 2 AGN Sample}\label{sec:sample}
\citet{reyes2008} presented a catalogue of 887 optically selected high luminosity Type 2 AGN with $z <$ 0.83, selected from the SDSS Data Release 6 \citep[DR6;][]{adelmanmccarthy2008}. The sample is selected using the following criteria; a redshift cut of $z < 0.83$ is applied so that \oiii{} is present in all spectra. A requirement for \oiii{} rest-frame equivalent width (EW) to be greater than 4~\AA{} is then employed to select objects with clear emission lines. A signal-to-noise ratio (S/N) cut of S/N $\geq$ 7.5 is performed with a metric that neglects continuum dominated sources with low S/N \citep[see][]{zakamska2003}. An \oiii{} line luminosity cut of \loiii{} $\geq 10^{41.9}\ $ergs$^{-1}$, which corresponds to an approximate bolometric luminosity cut of $L_{bol} \gtrsim 10^{44}\ $ergs$^{-1}$ \citep{lamastra2009}, is then applied to ensure completeness in emission line ratio selection. This procedure is followed by applying emission line ratio criteria to distinguish between a stellar and AGN ionizing continuum (see section \ref{sec:bpt}), constraining the sample to sources which satisfy either

\begin{equation}
 \log{\left(\frac{[\mathrm{O} \textsc{iii}]\lambda5007}{\mathrm{H}\beta}\right)} > \frac{0.61}{\log{([\mathrm{N}  \textsc{ii}]\lambda6583 / \mathrm{H}\alpha)} - 0.47} + 1.19
\end{equation}

or

\begin{equation}
 \log{\left(\frac{[\mathrm{O} \textsc{iii}]\lambda5007}{\mathrm{H}\beta}\right)} > \frac{0.72}{\log{([\mathrm{S} \textsc{ii}]\lambda\lambda6716,6731 / \mathrm{H}\alpha)} - 0.32} + 1.30
\end{equation}

for objects with $z < 0.36$. In the redshift range $0.36 \leq z < 0.83$ objects with a H$\beta$ detection with S/N > 3 are identified as an AGN candidate using the criterion

\begin{equation}
 \log{\left(\frac{[\mathrm{O} \textsc{iii}]\lambda5007}{\mathrm{H}\beta}\right)} > 0.3.
\end{equation}

Alternatively in this redshift range, if H$\beta$ is undetected while \oiii{} is detected a source is accepted as an AGN candidate. Finally each spectrum is visually inspected to reject objects exhibiting broad permitted emission lines and star-forming galaxies from the sample.

The selection criteria and visual inspection of each spectrum ensures that the sample contains the brightest Type 2 AGN with visible \oiii{} emission lines, with the authors estimating that 90$\%$ of luminous Type 2 AGN in the SDSS DR6 database are successfully selected.

\begin{figure}
 \includegraphics[width=\columnwidth]{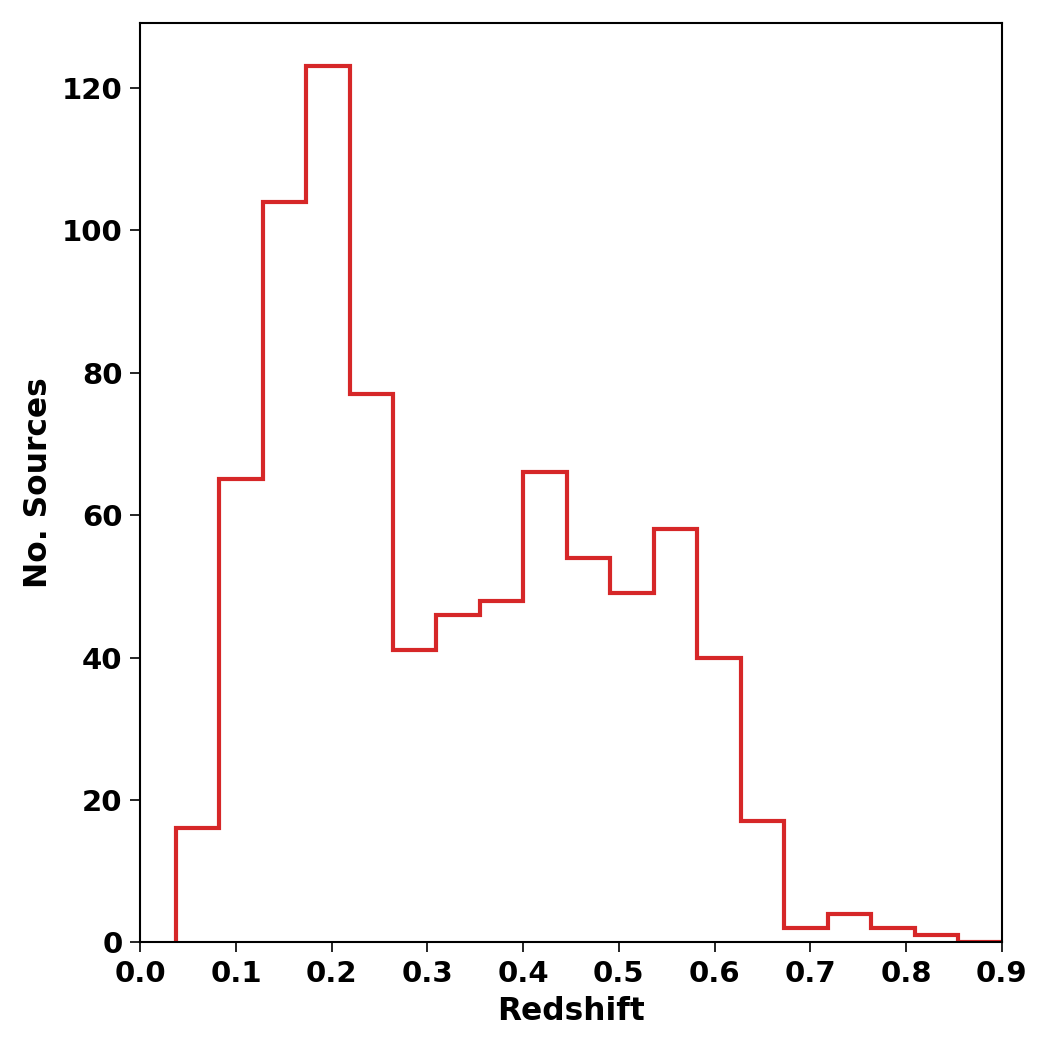}
 \caption{Distribution of redshifts for our sample of 813 sources derived from \citet{reyes2008} sample of optically selected luminous Type 2 AGN from SDSS DR6.}
 \label{fig:reyeszhist}
\end{figure}

We use our \qsfit{} Type 2 AGN recipe to analyse the \citet{reyes2008} sample of 887 Type 2 AGN. In a few cases we discarded spectra since the number of reliable spectral channels (SDSS quality mask = 0) amount to less than 50$\%$ of the total. We neglected 39 spectra due to this. We remove a further 34 spectra which have insufficient `good' spectral channels to successfully constrain the continuum components of our model. In addition to these omissions we were unable to locate one spectrum included in the \citet{reyes2008} sample, spec-0190-51457-0198, on the SDSS archive servers. We therefore have 813 sources in our sample appropriate to be analysed with \qsfit{}. Fig. \ref{fig:reyeszhist} shows the distribution of redshifts for the sources in our sample, which has a median redshift of z = 0.28.

All spectra are corrected for reddening using E(B-V) values derived using Schlegel, Finkbeiner and David galactic dust maps \citep{schlegel1998} with a Milky Way extinction curve \citep{odonnell1994} prior to fitting. We account for instrumental broadening assuming a Gaussian line spread function and using the SDSS spectral resolving power R$\sim$2000 to derive FWHM$_{instr}\sim$150$\ $kms$^{-1}$.

\section{QSFit: Type 2 AGN Fitting Recipe}\label{sec:qsfit_t2}

\begin{figure*}
\includegraphics[height=0.35\paperheight,width=\textwidth,keepaspectratio]{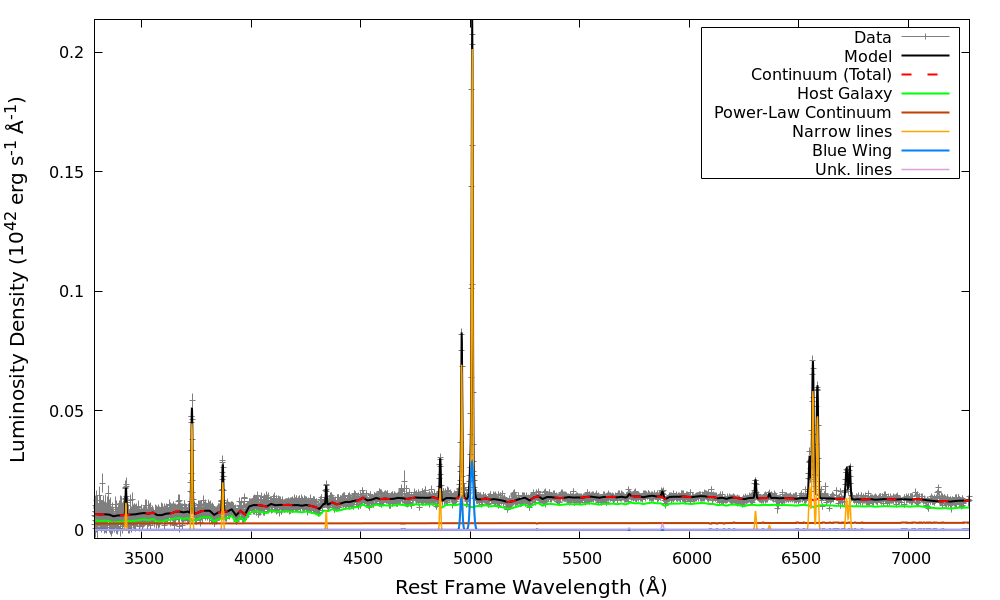}
\vfill
\includegraphics[height=0.27\paperheight,width=0.49\textwidth,keepaspectratio]{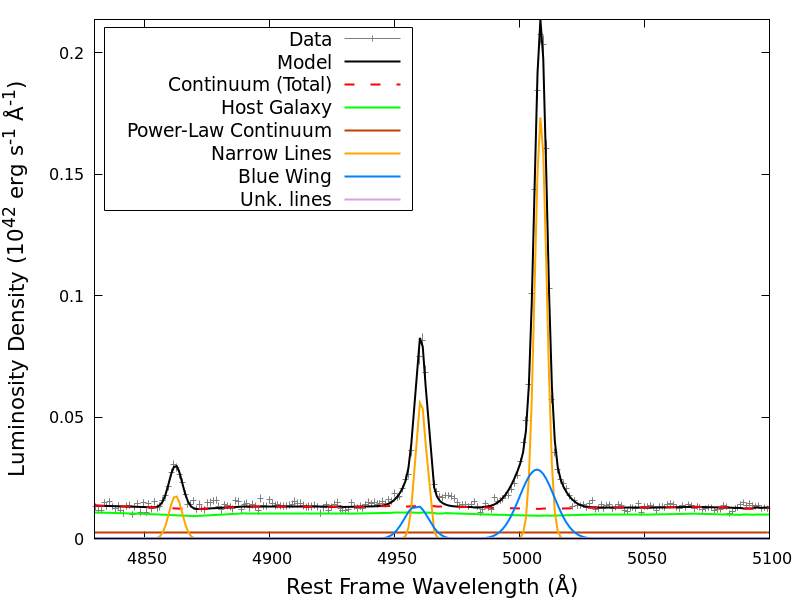}
\hfill
\includegraphics[height=0.27\paperheight,width=0.49\textwidth,keepaspectratio]{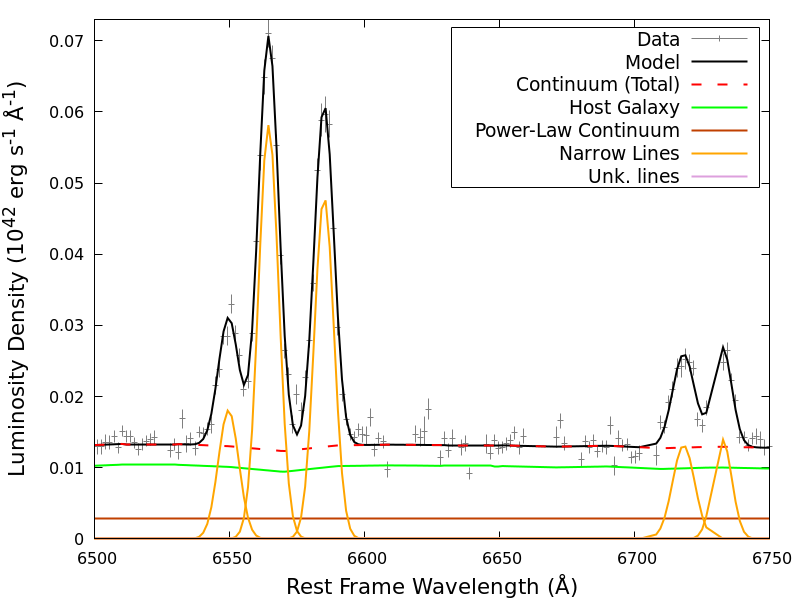}

\caption{Example spectral decomposition of an SDSS Type 2 AGN spectrum (SDSS J085205.38+024310.9) using \qsfit{} Type 2 AGN fitting routine developed in this work. The top panel shows the decomposition of the entire spectrum, while the bottom left and right panels present zoomed-in views of the [O {\sc iii}]+H$\beta$ and H$\alpha$+[N {\sc ii}] complex fits, respectively. In all three figures the solid black line represents the overall model attained by our fitting, the host-galaxy template is shown in solid green and the power-law AGN continuum in solid dark orange. The dashed red line presents the overall continuum level given by the sum of the host-galaxy and AGN continuum, while the solid orange lines represent the known narrow emission lines considered in the fitting recipe. The solid blue lines depict the [{\sc O iii}] blue wing asymmetry components and the solid purple lines show the unknown lines.}
\label{fig:t2_recipe}
\end{figure*}

In this section, we describe the models utilized for the primary spectral components used in our \qsfit{} Type 2 AGN recipe. The primary difference between optical Type 1 and Type 2 AGN is the presence of a dusty obscuring medium between the observer and the AGN central engine. In the optical spectra of AGN the obscuring medium manifests as a lack of components that physically originate within a few parsecs of the central engine. Table \ref{tab:recipediff} provides a summary of the components considered for Type 2 AGN as a comparison to those used in the default \qsfit{} recipe for Type 1 AGN (Calderone et al., in prep.). The Type 2 recipe does not include Balmer continuum and pseudo-continuum modelling or UV and optical iron complex templates. Fig. \ref{fig:t2_recipe} provides an example of an SDSS Type 2 AGN spectrum fit with our \qsfit{} Type 2 AGN recipe.

\begin{table}
 \caption{Main components included in the \qsfit{} Type 1 AGN (Calderone et al., in prep.) and Type 2 AGN fitting recipes.}
 \label{tab:recipediff}
 \centering
 \begin{tabular}{lcc}
  \hline
  Model Component & Type 1 & Type 2 \\
  \hline
  AGN Continuum & Single PL & Single PL\\
  Balmer Continuum (+ Pseudo-Continuum) & Yes & - \\
  Host-Galaxy Template & Optional & Yes \\ 
  UV and Optical Iron Templates & Yes & - \\
  Known Emission Lines & Yes & Yes \\
  Known Emission Line Profiles & Gaussian & Gaussian \\ 
  Unknown Lines & 10 & 2 \\
    
  \hline
 \end{tabular}
\end{table}

\subsection{AGN Continuum}\label{sec:continuum}
The majority of the AGN continuum emission emanating from the central accretion disk is obscured in Type 2 AGN. Many sources, however, still require the presence of a detectable base continuum that becomes increasingly subdominant to host-galaxy contributions as the AGN luminosity decreases. 

\qsfit{}'s Type 2 AGN recipe employs a single power-law to model the continuum of the form:

\begin{equation}\label{eqn:pl}
 L_{\lambda} = A \left( \frac{\lambda}{\lambda_{\rm 0}} \right) ^{\alpha_{\rm 0}}
\end{equation}

where $\lambda_{0}$ is a reference wavelength, $A$ is the luminosity density at $\lambda = \lambda_{0}$ and $\alpha_{\rm 0}$ is the spectral slope. We fix $\lambda_{0}$ at the median value of the available spectral sample wavelengths, $A$ is constrained to be positive and $\alpha_{\rm 0}$ is constrained to be in the range [-5, 5].

\subsection{Host-Galaxy Template}
The obscured nature of Type 2 AGN results in non-negligible host-galaxy contributions to their optical spectra being commonplace, particularly in the regime of low luminosity AGN. This factor necessitates effective modelling of host-galaxy light for the analysis of Type 2 AGN spectra. 

\qsfit{} incorporates a library of host-galaxy SEDs, consisting of the SWIRE templates collated by \citet{polletta2007} supplemented with twelve BC03 \citep[][]{bruzal&charlot2003} stellar population synthesis model generated templates used to model COSMOS galaxies by \citet{ilbert2009}. The SWIRE library contains 25 SEDs, generated using the GRASIL code \citep{silva1998}, ranging from quiescent Elliptical galaxies through to Starbursts and Type 1 AGN. The supplementing \citet{ilbert2009} library of 12 BC03 templates consists of starburst templates with ages ranging from 0 to 11$\ $Gyr. For the purpose of modelling AGN hosts, templates including AGN emission are disregarded from consideration as \qsfit{} deals with the AGN component itself, leaving a total of 27 elliptical, spiral and starburst templates from the parent library.

Despite this wide range of available host-galaxy templates, we have found that the 5$\ $Gyr elliptical template from the SWIRE library is appropriate and effective when modelling the continuum emission of the Type 2 AGN in our sample. Consequently, we adopt this as our default host-galaxy in the following analysis of Type 2 AGN spectra. The only free parameter in the model associated to the host-galaxy component is its normalization (luminosity) at 5500\AA{}.

\subsection{Known Emission Lines}\label{sec:knownemlines}
\qsfit{} incorporates a spectral transition database of common permitted, forbidden and intercombination AGN atomic transitions collected from The Atomic Line List\footnote{\url{https://www.pa.uky.edu/~peter/atomic/}}. All emission lines in \qsfit{} are considered using their vacuum wavelengths. Known emission lines represent a subset of these spectral transitions that are considered in a given fitting recipe. The known emission lines considered for our Type 2 AGN recipe and their wavelengths are given in Table \ref{tab:knownemlines}. Each line considered by a recipe can be configured as "Broad", "Narrow", "VeryBroad", or a combination of these (see Calderone et al., in prep. for detailed definitions). Each line can also be assigned to have a Gaussian, Lorentzian or Voigt line profile. In this work all emission lines are modelled with a Gaussian line profile unless explicitly specified as otherwise. 

For Type 2 AGN the Narrow line profile is solely utilised and is modified with respect to the Type 1 AGN recipe to have FWHM constrained in the range [10 - 2,000]$\ $km\ s$^{-1}$. A reduced lower limit on the FWHM of narrow lines is required to appropriately fit the intrinsically narrower lines of Type 2 AGN when compared with Type 1. The lower limit to the FWHM is set by the expected scale of thermal broadening (i.e. ignoring bulk kinematics of any line), so any observed line will have a FWHM larger than this. This theoretical limit should never be met however, as the SDSS spectral resolution of 150$\ $kms$^{-1}$ limits the narrowest measurable line FWHM to this value.

A subset of the lines considered are members of emission line doublets (eg. [{\sc O ii}]$\lambda$3727). In cases where the individual emission lines of a doublet cannot be distinguished in a typical SDSS spectrum, the line complex is fit with a single narrow line component placed at the wavelength of the midpoint between the two line locations (a velocity offset term is used in the fit to allow varying this value to best fit the data).

\begin{table}
 \caption{List of emission lines considered in the Type 2 AGN \qsfit{} recipe and their rest-frame vacuum and air wavelengths.}
 \label{tab:knownemlines}
 \centering
 \begin{tabular}{lcc}
  \hline
  Line & Vacuum Wavelength (\AA) & Air Wavelength (\AA)\\
  \hline
  Mg {\sc ii} & 2796.35 & 2795.53 \\ [2pt]
  [Ne {\sc v}] & 3426.50 & 3425.88 \\ [2pt]
  [{\sc O ii}] & 3728.48 & 3727.42 \\ [2pt]
  [Ne {\sc iii}] & 3870.16 & 3868.76 \\ [2pt]
  H$\gamma$ & 4341.68 & 4340.47 \\ [2pt]
  H$\beta$ & 4862.68 & 4861.33 \\ [2pt]
  [{\sc O iii}] & 4960.30 & 4958.91 \\ [2pt]
  [{\sc O iii}]bw & 4960.30 & 4958.91 \\ [2pt]
  [{\sc O iii}] & 5008.24 & 5006.84 \\ [2pt]
  [{\sc O iii}]bw & 5008.24 & 5006.84 \\ [2pt]
  [{\sc O i}] & 6302.05 & 6300.30 \\ [2pt]
  [{\sc O i}] & 6365.54 & 6363.78 \\ [2pt]
  [{\sc N ii}] & 6549.85 & 6548.05 \\ [2pt]
  H$\alpha$ & 6564.61 & 6562.82 \\ [2pt]
  [{\sc N ii}] & 6585.28 & 6583.46 \\ [2pt]
  [{\sc S ii}] & 6718.29 & 6716.44 \\ [2pt]
  [{\sc S ii}] & 6732.67 & 6730.81 \\ [2pt]
  \hline
 \end{tabular}
\end{table}

The \oiii{} emission line is among the most ubiquitous and brightest emission lines observed in AGN and Star-forming Galaxy (SFG) optical spectra. Due to its prevalence and high luminosity, the measured properties of the \oiii{} line are often significantly affected by the kinematics of the NLR. Bulk motions produce distortions such as blue wings when a portion of NLR gas is moving radially towards the observer, which are often used as tracers for outflows in AGN \citep[eg.][]{zakamska2014}. Our Type 2 AGN recipe includes an \oiii{} asymmetric tail component to model the line asymmetries of \oiii{} lines where it is required ([O {\sc iii}]bw in Table \ref{tab:knownemlines}). This emission line feature is modelled with a Gaussian narrow line component constrained to have a velocity offset on the blue (shorter wavelength) side of its parent line as well as FWHM larger than its parent line to ensure it is used to model asymmetries at the base of the emission line as intended. 

Due to quantum mechanical transition probabilities the [O {\sc iii}]$\lambda$4959 line profile exhibits an identical form to that of \oiii{} with a third of the luminosity (see Sect. \ref{sec:oiii_flux_ratio}). We therefore include a blue wing component for the [O {\sc iii}]$\lambda$4959. \qsfit{} allows for two parameters to be fit using a single free parameter value, reducing the number of free parameters in the model. In these cases the two parameters are ``patched" to one another. We use the patching feature of \qsfit{} to provide a simultaneous fit for the \oiii{} and [O {\sc iii}]$\lambda$4959 components where we constrain the luminosity of the [O {\sc iii}]$\lambda$4959 blue wing to be one third times the luminosity of the \oiii{} blue wing component and to share the same FWHM and velocity offsets. This system allows the correct asymmetric profile component of [O {\sc iii}]$\lambda$4959 to be modelled even in spectra where the asymmetry is poorly constrained due to the S/N of the spectrum.

\subsection{Unknown Lines}
In the final stages of \qsfit{}'s spectral fitting analysis, Gaussian-profile ``unknown'' lines are placed at areas of highest residuals with the aim to model emission lines that are present in the spectrum but have not been included in the known spectral lines library of the current recipe. 

As Type 1 AGN optical spectra contain broader BLR components, some of which are not seen in the NLR, Type 2 AGN inherently require fewer components to model. Due to the reduced degrees of freedom it is important that \qsfit{}'s unknown lines are not relied upon to fill large areas of residuals in Type 2 AGN spectra, which occurs when there are more unknown lines considered than there are genuine unaccounted-for emission lines in the spectrum. To this end, in our Type 2 AGN recipe the number of unknown lines is reduced from the default value of ten to two. This value may need to be adjusted according to the spectral coverage and redshift of a given spectrum or sample of spectra and is easily done so through \qsfit{}'s \verb|source.options| dictionary. Using two unknown lines is found to give qualitatively good fits for SDSS Type 2 AGN spectra analysed in this work (z $\leq$ 0.83). 

We constrain the FWHM of unknown lines to lie in the range [10 - 2,000]~kms$^{-1}$ reducing both the maximum and minimum values from the default constraints in the range [500 - 1$\times$10$^{4}$]~kms$^{-1}$. This further prevents unknown lines from being used to fill large areas of residuals, which cause the host-galaxy template to have an incorrect normalisation in some cases, and forces them into their intended role of being placed at locations of lines not considered in the current set of known spectral lines. We do not fix the FWHM of unknown lines to those of known emission lines because they may be from kinematically or spatially different components.

\section{Baseline Type 2 Recipe Results}\label{sec:sample_results}

In this section we present the results of the baseline fitting recipe described in Sect. \ref{sec:qsfit_t2}. We discuss spectral properties derived from our sample of 813 luminous Type 2 AGN spectra analysed with our Type 2 AGN \qsfit{} recipe. Using multiple emission line diagnostics we check the self-consistency and consistency with the literature of our results. We publish a catalogue of our luminous Type 2 AGN sample spectral measurements alongside this paper (see App. \ref{app:cat_descr} for a description and download instructions).

\subsection{Line Profiles}\label{sec:gvslprofiles}
Gaussian and Lorentzian emission line profiles are the most widely accepted for the modelling of AGN emission lines. The literature is undecided on which profile is most effective to use; some works employ Gaussian profiles \citep[eg.][]{sexton2021, sarzi2006, kuraszkiewicz2004} and others Lorentzian profiles, particularly for narrow line Seyfert 1 (NLS1) and highly accreting AGN \citep[eg.][]{marziani2013, sulentic2002, veroncetty2001}. Alternatively, some works opt to use both profiles and keep the result with the best fit statistic \citep[eg.][]{diasdossantos2022, shen2011, liu2010}. For the majority of spectral resolutions and S/N the two line profiles are largely interchangeable. Some works explore the underlying line profile of AGN emission lines with both \citet{kollatshny2013} and \citet{naddaf2021} reporting that broad AGN emission lines intrinsically tend towards a Lorentzian profile. \citet{naddaf2021} also highlight the non-triviality of emission line profile generation with the overall profiles depending on a host of parameters from viewing angle to dust-to-gas mass ratio. \citet{berton2019} report that some NLS1 galaxies from the SDSS have narrow emission lines best modelled using Gaussian profiles while others with Lorentzian profiles. They show the two samples present different physical and energetic properties, suggesting that line profiles may evolve with the AGN over time.

We perform \qsfit{} analysis of our sample using both Gaussian and Lorentzian line profiles for all emission lines considered in our recipe. Gaussian line profiles give a median $\chi^{2}_{red}$ value of 1.39$\pm$0.36 across the sample, while Lorentzian line profiles give a median of 1.45$\pm$0.38. Given the insignificant change in $\chi^{2}_{red}$ between the two line profiles, there is no clear evidence to choose one over the other. We therefore opt to use Gaussian profiles for this work. This is not to say that the underlying profile is Gaussian, but that this provides the best results for the instrumental resolution and S/N of low-z SDSS spectra. Indeed, we explore in Sect. \ref{sec:snratio} with high S/N emission lines that the underlying profiles are non-Gaussian and require multiple components to be correctly modelled. 

\subsection{Continuum Slope}
\begin{figure}
 \includegraphics[width=\columnwidth]{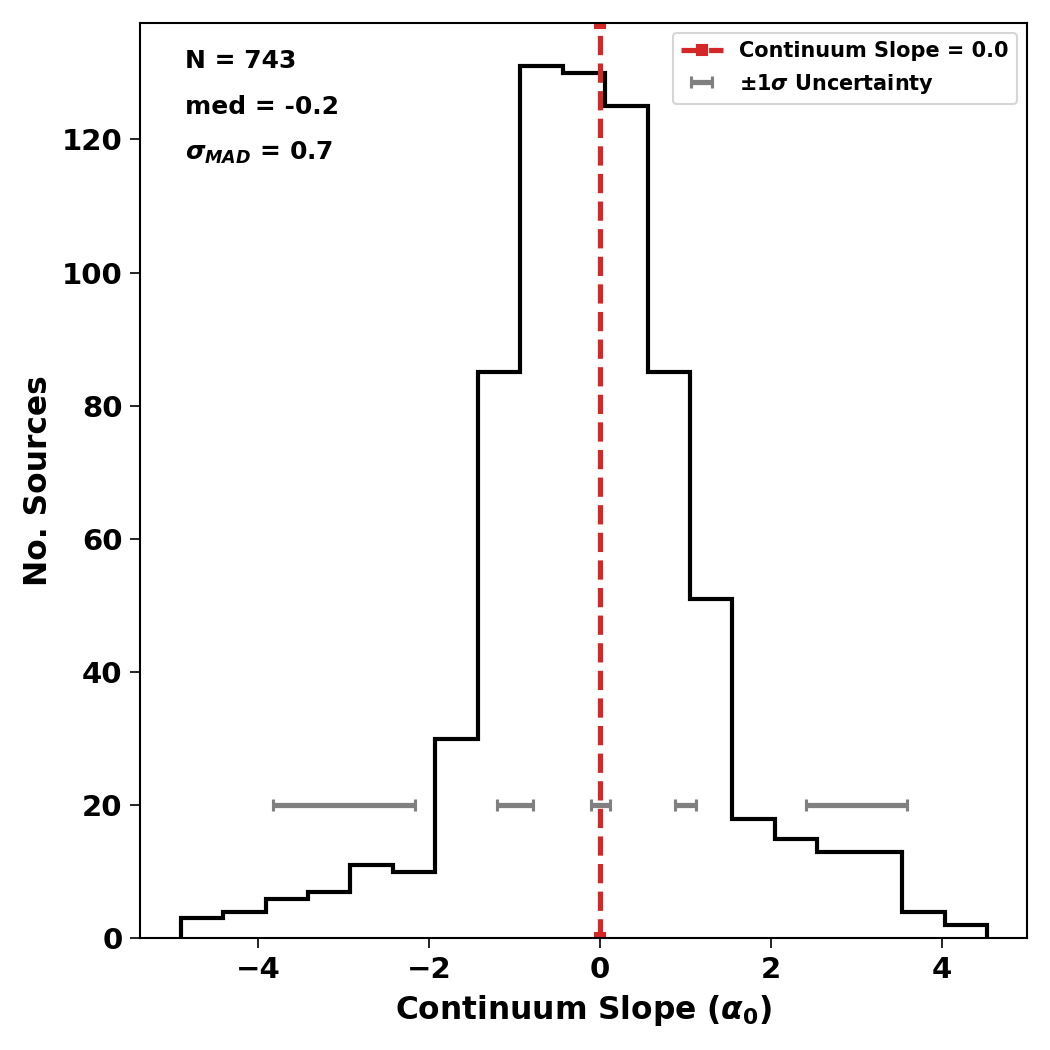}
 \caption{Distribution of power-law continuum spectral slopes ($\alpha_{\rm 0}$) for 743 luminous Type 2 AGN. The red line is drawn at a continuum slope value of 0, which corresponds to a flat continuum. Grey bars depict representative $\pm$1$\sigma$ statistical uncertainties.}
 \label{fig:continuum_slope}
\end{figure}

The distribution of power-law continua spectral slopes ($\alpha_{\rm 0}$, in equation \ref{eqn:pl}) for our sample along with representative error bars for different measured $\alpha_{\rm 0}$ values is shown in Fig. \ref{fig:continuum_slope}. The distribution shows that in cases where a continuum component is required an approximately even spread of blue continua ($\alpha_{\rm 0}$ < 0) and red continua ($\alpha_{\rm 0}$ > 0) are utilized. Outliers beyond $\alpha_{\rm 0}$ = $\pm$2 have large statistical and un-quantified systematic uncertainties with a median 1$\sigma$ statistical uncertainty of 0.3 for these values of $\alpha_{\rm 0}$. Measurements of $\alpha_{\rm 0}$ within this range have much lower statistical uncertainties with a median 1$\sigma$ uncertainty of 0.07. This indicates that the true distribution is likely to lie within the $\alpha_{\rm 0}$ = $\pm$2 range in continuum slope values and is broadened by the associated uncertainties.

In works analysing the power-law continuum slopes of Type 1 AGN steeper negative values are found. \citet{calderone2017} report an average slope of $\alpha_{\rm 0}$ = -1.7 for Type 1 AGN with z $\sim$ 0.7. The median continuum slope value of our Type 2 AGN sample is $\alpha_{\rm 0}$ = -0.2$\pm$0.7. The difference between our median value for Type 2 AGN and that found by \citet{calderone2017} for Type 1 AGN may well indicate the effect of reddening and obscuration on the underlying continuum on Type 2 sources. However, some caution should be exercised with this interpretation as it is also possible that the result is in part due to the power-law filling residuals that that arise from modelling the stellar population with a single host-galaxy template. 

\subsection{BPT Diagrams}\label{sec:bpt}
Named after their pioneers Baldwin, Phillips and Terlevich \citep{bpt1981}, BPT diagrams are a set of emission line diagnostic plots used to separate AGN (referred to as Seyfert in the diagrams) from pure star forming galaxies (also referred to as HII galaxies) and LINERs (Low-Ionization Nebula Emission Regions). These non-Seyfert classes possess similar spectroscopic properties and are more numerous than AGN. By comparing line flux ratios of narrow emission lines which have varying degrees of ionization, the parameter space of the BPT diagrams can be divided into regions that discern the nature of the ionizing continuum in the object (ie. from OB type stars; star formation, power-law AGN continua or interstellar medium shocks).

The most prominent of these diagnostic diagrams compare the ratio of \oiii{}/H$\beta$ lines with the corresponding ratios of other pairs of neighbouring lines, such as [N {\sc ii}]$\lambda$6583/H$\alpha$, [O {\sc i}]$\lambda$6300/H$\alpha$ and [S {\sc ii}]$\lambda$6716$,\lambda$6731/H$\alpha$. Each of these diagrams is supplemented with a theoretical maximum starburst line, introduced by \cite{kewley2001} who used stellar population synthesis and photoionization models to determine the emission line ratio upper limits for pure stellar photoionization models. The [N {\sc ii}]$\lambda$6583/H$\alpha$ diagram makes use of a modification of the \cite{kewley2001} starburst line, defined by \cite{kauffmann2003}, which empirically divides pure star-forming galaxies from AGN-HII composite sources. \cite{kewley2006} derive an empirical boundary used to cleanly separate AGN (Seyferts) and LINER sources in the [O {\sc i}]$\lambda$6300/H$\alpha$ and [S {\sc ii}]$\lambda$6716$\lambda$6731/H$\alpha$ diagrams.

\begin{figure*}
\centering
\includegraphics[width=.33\textwidth]{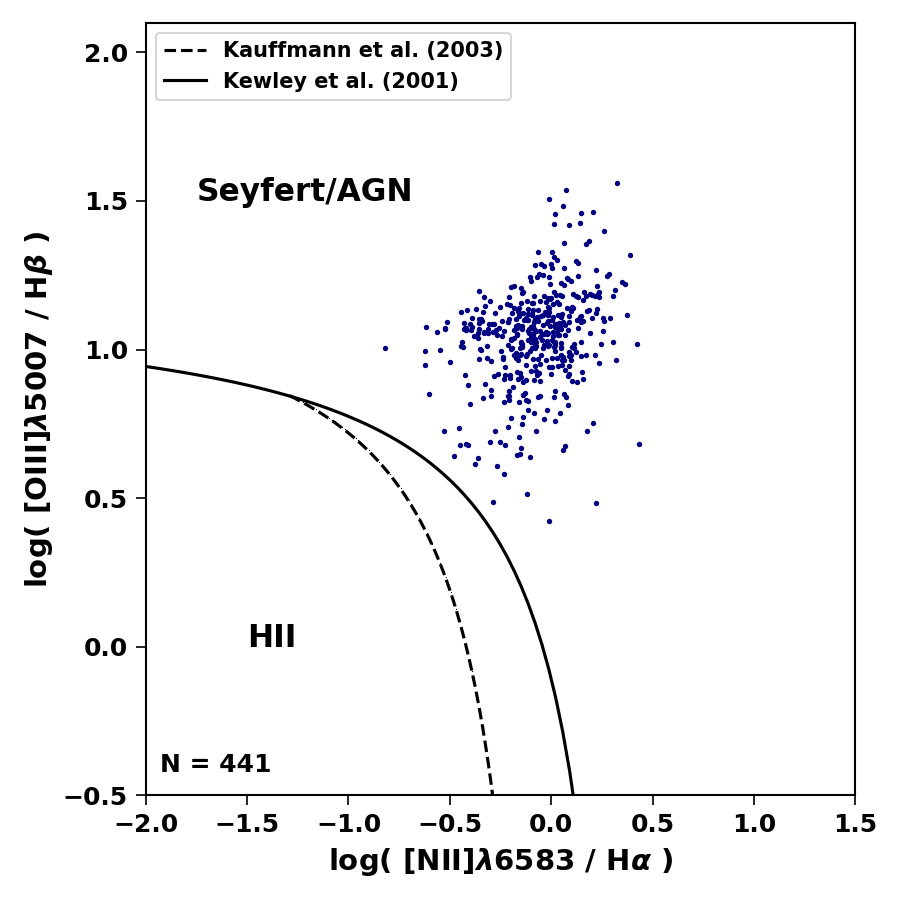}\hfill
\includegraphics[width=.33\textwidth]{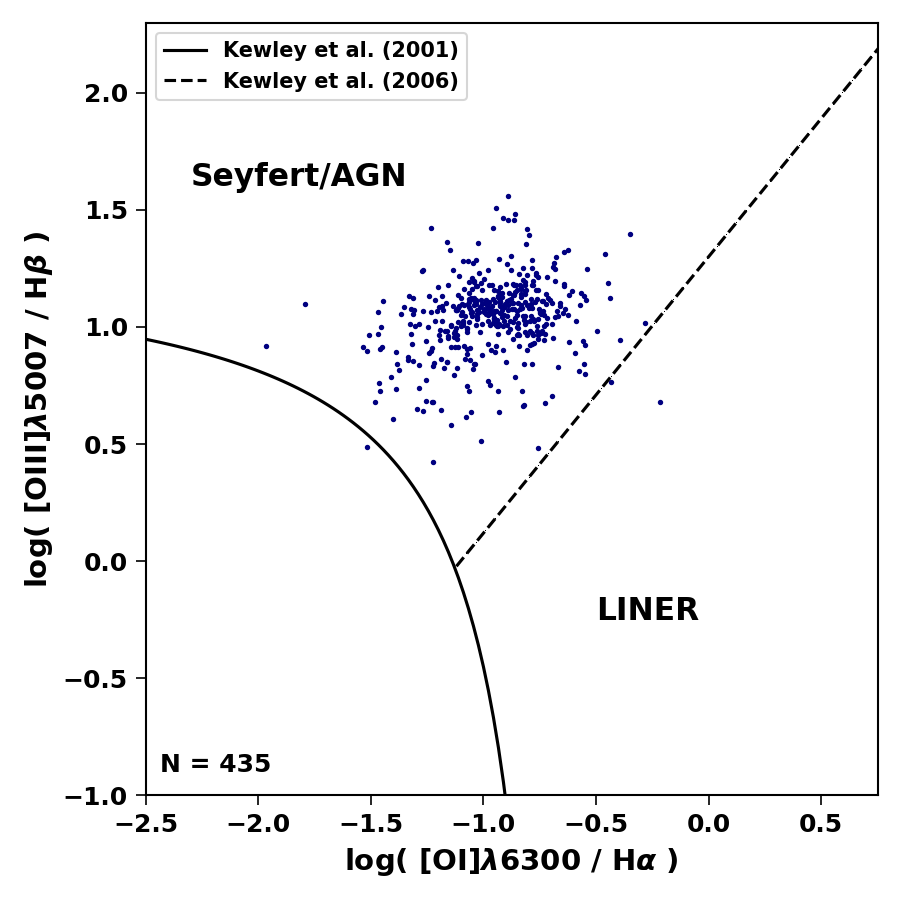}\hfill
\includegraphics[width=.33\textwidth]{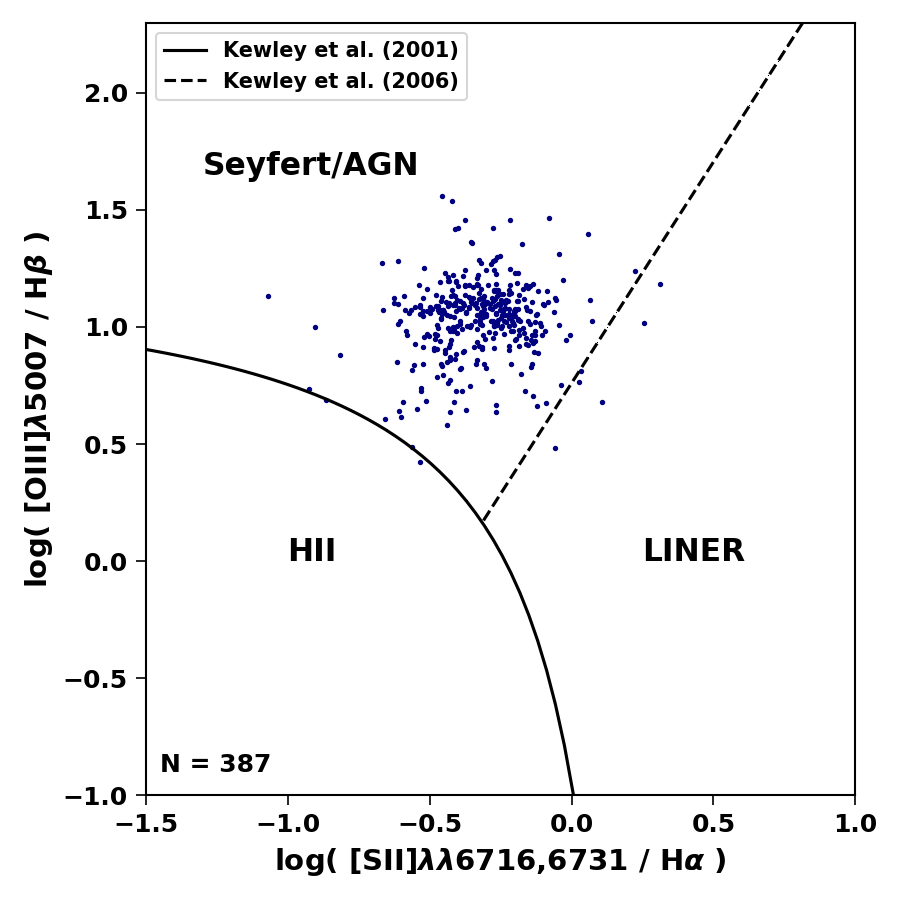}
\caption{BPT line diagnostic diagrams plotted for \qsfit{} emission line measurements of Type 2 AGN, for which the four emission lines required for each diagram are considered in the fit and well constrained. The solid black curve in all three diagrams shows the theoretical maximum starburst line dividing AGN from HII regions (star-forming galaxies) \citep{kewley2001}. (\textit{left}) \oiii{}/ H$\beta$ vs [N {\sc ii}]$\lambda$6583/H$\alpha$ diagram. The black dashed curve defines the empirical AGN-HII region demarcation \citep{kauffmann2003}. (\textit{centre}) \oiii{}/H$\beta$ vs [O {\sc i}]$\lambda$6300/H$\alpha$ diagram. The dashed black line represents the empirical separation between Seyfert (AGN) and LINER driven sources \citep{kewley2006}. (\textit{right}) \oiii{}/H$\beta$ vs [S {\sc ii}]$\lambda\lambda$6716,6731/H$\alpha$ diagram. The dashed black line holds the same meaning as the center plot.}
\label{fig:bpt}
\end{figure*}

Narrow emission line ratios as measured in our analysis are plotted on BPT diagrams as a self-consistency check of the measurements. The three BPT diagrams discussed above are plotted in Fig. \ref{fig:bpt} for the sources in our analysed sample where all four emission lines needed for each diagram are available in the spectrum. As expected from the \citet{reyes2008} \oiii{} line luminosity selection criteria, all the sources exhibit large values of \oiii{}/H$\beta$ which to first order traces ionization parameter in the Seyfert/LINER branch. This places them in the top portion of each diagram. Every source analysed with \qsfit{} falls into the AGN region of the [N {\sc ii}]$\lambda$6583/H$\alpha$ diagram (fig. \ref{fig:bpt}a). This is the expected result as \citet{reyes2008} makes use of emission line ratio cuts based on the \citet{kewley2001} maximum starburst line in [N {\sc ii}]$\lambda$6583/H$\alpha$ space for their sample selection. Thus, emission line measurements derived with our \qsfit{} Type 2 AGN recipe are compatible with expectations. 

\begin{figure}
\includegraphics[width=0.48\textwidth, keepaspectratio]{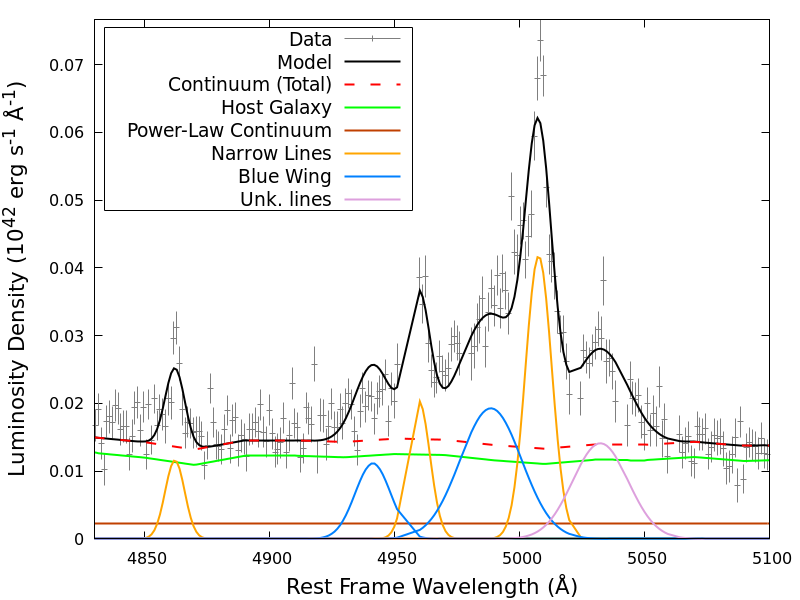}
\caption{Example case (SDSS J084314.55+285059.9) where a poor fit is achieved using our Type 2 AGN \qsfit{} recipe due to a heavily asymmetric [{\sc O iii}] complex. Component colour scheme identical to Fig. \ref{fig:t2_recipe}.}
\label{fig:strange_oiiicomplex}
\end{figure}

Sources that stray from the central bulk of points in the [N {\sc ii}]$\lambda$6583/H$\alpha$ diagram consist of spectra that exhibit strong asymmetries in the [O {\sc iii}]+H$\beta$ complex and spectra with double peaked emission lines, for which additional components are required to model in both cases. Figure \ref{fig:strange_oiiicomplex} presents an example of such an [O {\sc iii}]+H$\beta$ complex. Double peaked Type 2 AGN spectra are treated in section \ref{sec:dp}.

Both the [O {\sc i}]$\lambda$6300/H$\alpha$ (Fig. \ref{fig:bpt} middle panel) and [S {\sc ii}]$\lambda$6716$\lambda$6731/H$\alpha$ (Fig. \ref{fig:bpt} right panel) diagnostic diagrams have the vast majority of \citet{reyes2008} sources lie in the AGN portion of their plots. The former of these plots has three sources and the latter ten sources that stray into the HII galaxy or LINER region of the diagrams. Visual inspection of these non-AGN classified sources shows that the fits are acceptable and the sources genuinely lie in these portions of the diagram.

\subsection{[O {\sc iii}]$\lambda$5007: [O {\sc iii}]$\lambda$4959 Flux Ratio}\label{sec:oiii_flux_ratio}
In low density NLR-like environments in which collisional de-excitation is negligible, the relative number of \oiii{} and [O {\sc iii}]$\lambda$4959 photons is governed by the ratio of the probabilities of the quantum mechanical transitions that produce them. Hence, the ratio of the fluxes in the \oiii{} to [O {\sc iii}]$\lambda$4959 emission lines are fixed at a 3:1 ratio \citep{osterbrock1989}. Although the asymmetric blue wing components of these lines are fixed at a 3:1 luminosity ratio in our recipe, we do not constrain the core components of the [O {\sc iii}] emission lines to conform to a fixed ratio in our analysis. The fixed nature of this [O {\sc iii}] emission line ratio therefore provides a self-consistency check for \qsfit{}, as the observed flux ratio inherent to the spectra should be recovered by the core components in our fitting.

\begin{figure}
 \includegraphics[width=\columnwidth]{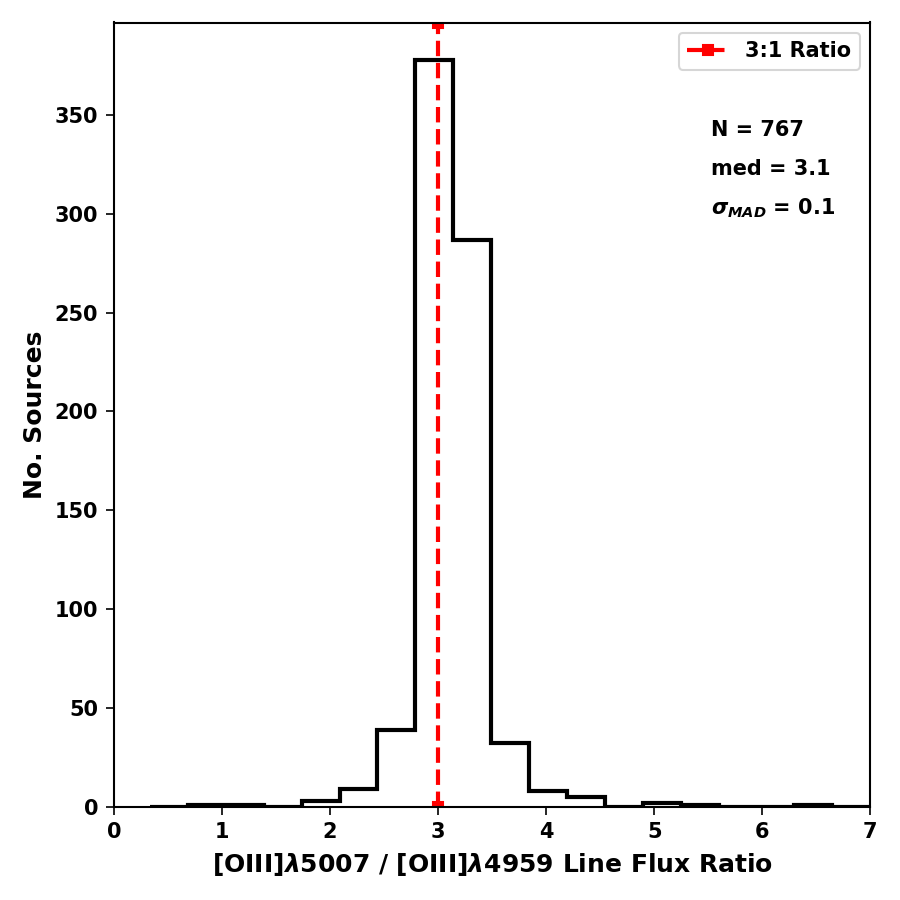}
 \caption{Distribution of \oiii{} /[O {\sc iii}]$\lambda$4959 line flux ratios for \qsfit{} measurements of Type 2 AGN for which both lines are considered in the fit and are well constrained. The red dashed line is drawn at a flux ratio of 3:1, the value predicted by quantum mechanical transition probabilities \citep{osterbrock1989}.}
 \label{fig:oiiiratiohist}
\end{figure}

Fig. \ref{fig:oiiiratiohist} gives a histogram of the ratio of \oiii{} to [O {\sc iii}]$\lambda$4959 emission line luminosities (sum of luminosity in core and blue wing components) as measured by \qsfit{} for the 767 sources in our sample for which the two lines available. Across the sample, we measure a ratio of 3.1$\pm$0.1. This measurement is slightly larger, although
compatible, with the value of 3 used in \citet{reyes2008} and found by \cite{liu2010}, while smaller than those reported by \cite{francis1991} and \cite{vandenberk2001} (3.66 and 3.63, respectively). In the latter cases, however, the [O {\sc iii}] ratio has been estimated on a composite quasar spectrum of a Type 1 AGN, while here we are providing the average of ratios as measured in individual Type 2 AGN spectra.

The Type 2 spectra in our analysis which deviate from the modal value observed in Fig. \ref{fig:oiiiratiohist} are exclusively comprised of low S/N spectra. Many of these outliers have masked spectral channels in the [O {\sc iii}] complex region or features not accounted for by our model, such as double peaked emission lines treated in Sect. \ref{sec:dp}.

\subsection{Balmer Decrement}
Hydrogen Balmer lines are permitted Hydrogen recombination transitions from higher orders to the $n=2$ state and are among the most prominent lines observed in AGN spectra. The brightest of the Balmer lines in low-z AGN are typically H$\alpha$ and H$\beta$, produced from electrons originating in the $n=3$ and $n=4$ state, respectively. The Balmer decrement ratio H$\alpha$/H$\beta$ exhibits a constant value over all sources with similar line emitting gas conditions. Using optically thick recombination, an intrinsic H$\alpha$/H$\beta$ value of 2.87 is predicted for HII regions photoionized by a hot star, while a value of 3.1 is predicted for AGN NLRs \citep{osterbrock1989}. The value of H$\alpha$/H$\beta$ is elevated for NLRs of AGN due to H$\alpha$ emission being enhanced by collisional excitation from higher density gas and the creation of a partly ionized transition region by the AGN ionizing continuum, which is much harder than those produced by stars \citep{gaskell&ferland1984, halpern&steiner1983}.

\begin{figure}
 \includegraphics[width=\columnwidth]{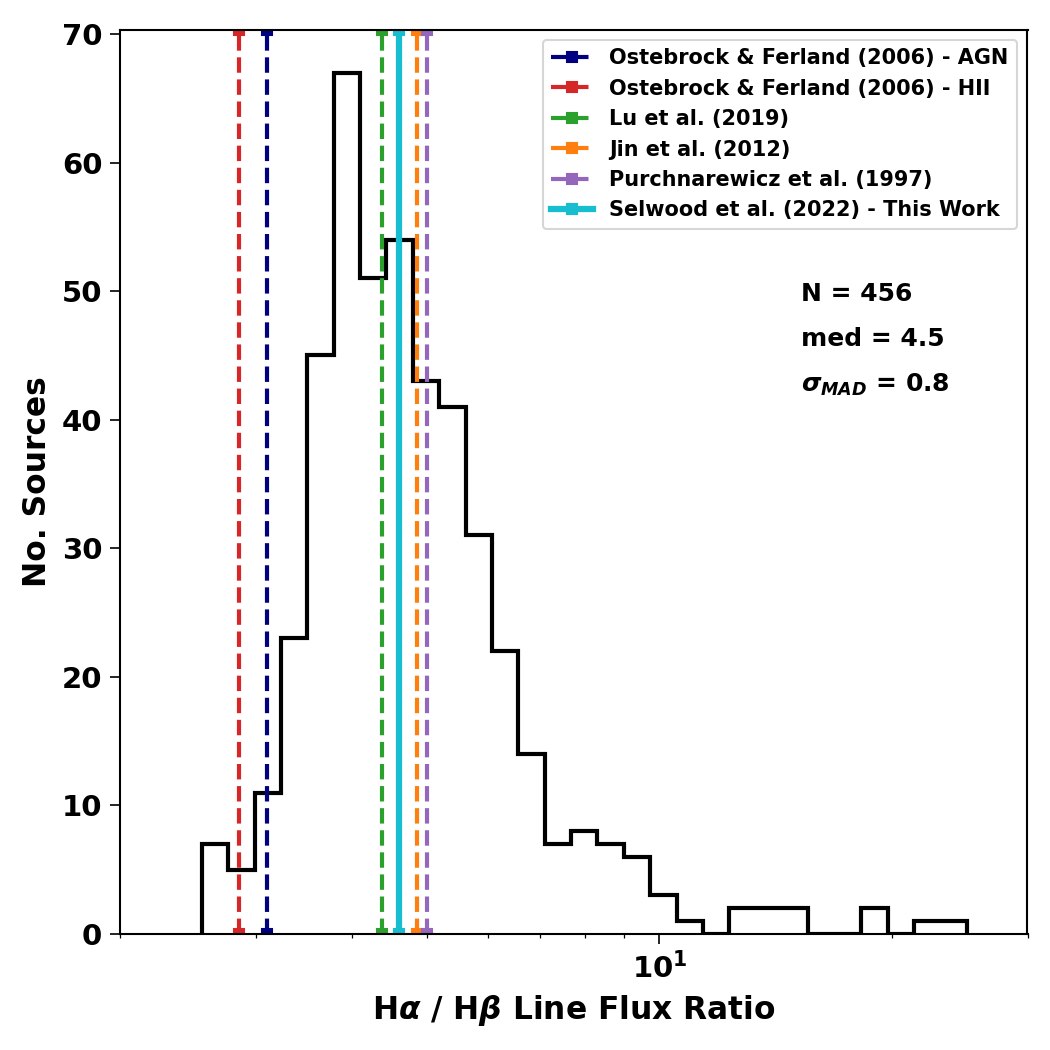}
 \caption{Distribution of Balmer decrement values (H$\alpha$/H$\beta$ emission line flux ratio) for 456 sources in which both lines are available in the spectrum. The solid cyan line denotes the median value measured in this work. Theoretically derived Balmer decrement values for gas photoionized by a hot star (HII region) and gas photoionized by an AGN continuum are shown with dashed red and blue lines, respectively \citep{osterbrock1989, halpern&steiner1983}. Narrow line Balmer decrements derived by other works are shown with dashed lines; \citet{lu2019} (\textit{green}), \citet{jin2012} (\textit{orange}) and \citet{purchnarewicz1997} (\textit{purple}).}
 \label{fig:balmerdec}
\end{figure}

In Fig. \ref{fig:balmerdec} we show the H$\alpha$/H$\beta$ Balmer decrement measured for the 456 sources in our sample where both emission lines are available in the spectrum. The dashed red line represents the theoretical optically thick recombination value of the H$\alpha$/H$\beta$ decrement for HII regions, while the dashed blue line represents the same for AGN NLRs. The median value of H$\alpha$/H$\beta$ in our sample is 4.5$\pm$0.8 (solid cyan line in Fig. \ref{fig:balmerdec}). The discrepancy in expected theoretical and measured values of the Balmer decrement is postulated to be due to intrinsic dust extinction from obscuring dust at the redshift of the AGN, which is unaccounted for in our analysis. Line of sight dust extinction will absorb, and therefore diminish emission preferentially at shorter wavelengths. H$\alpha$ and H$\beta$ are separated by a relatively large wavelength span, with H$\beta$ observed at a shorter rest-frame wavelength of 4863\AA{} compared to H$\alpha$ at 6564\AA{}. Therefore intrinsic reddening by dust at the redshift of a source will act to reduce the flux observed from the H$\beta$ emission line, while having a much weaker effect on the flux from the H$\alpha$ emission line. The effect of this is to skew the H$\alpha$/H$\beta$ Balmer decrement to a higher value than would be measured if intrinsic dust reddening was accounted for using a dust reddening curve or not present. Through comparison of observed and theoretically predicted values, the Balmer decrement is often utilized as a means to measure the intrinsic obscuration towards a line emitting region \citep{gaskell2017, dominguez2013}. Adopting this methodology implies an average NLR optical reddening in our sample of $A_{V}\approx$ 1$\ $mag.

\begin{figure}
 \includegraphics[width=\columnwidth]{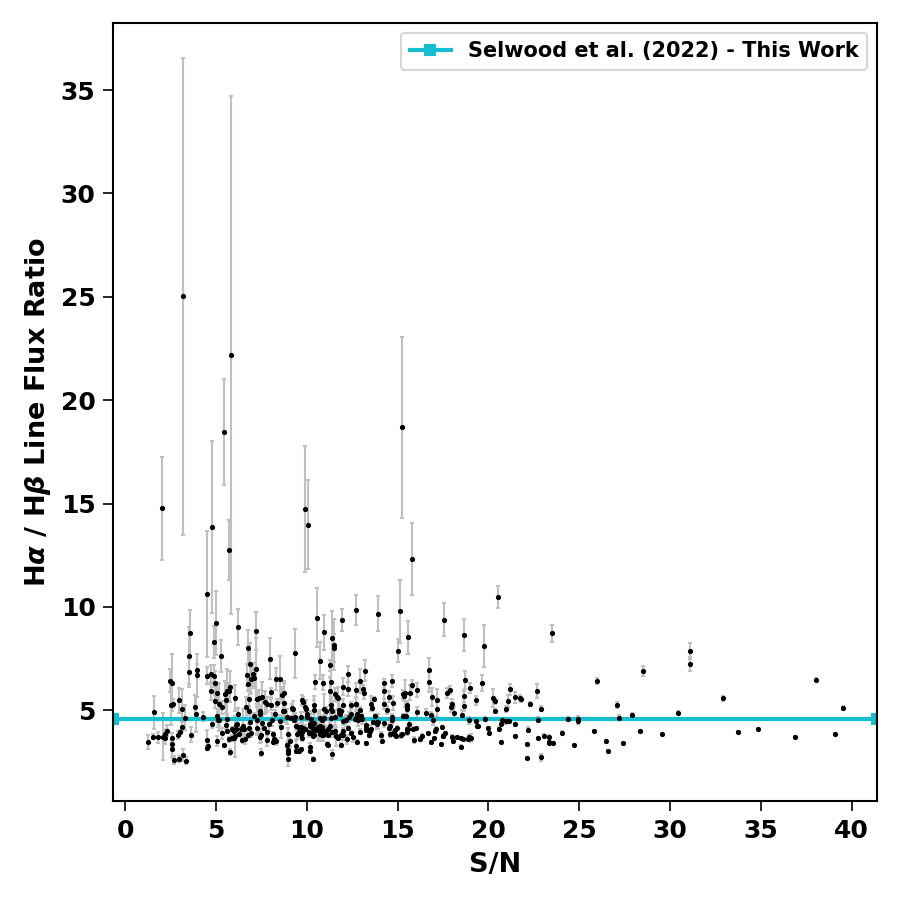}
 \caption{Measured Balmer decrement values for our sample of luminous Type 2 AGN as a function of median spectrum S/N. The solid cyan line represents our measured median Balmer decrement value of 4.5 for our sample of luminous Type 2 AGN spectra. Grey error bars corresponding to the $\pm$1$\sigma$ statistical uncertainties of our measurements are visualised.}
 \label{fig:balmerdec_unc}
\end{figure}

The S/N of each spectrum in our sample is estimated by dividing the median flux value of the spectrum by the median uncertainty. This metric provides an estimate of the continuum S/N for each spectrum as the majority of spectral pixels consist of continuum emission. Fig. \ref{fig:balmerdec_unc} presents our Balmer decrement measurements as a function of median S/N with their associated statistical uncertainties visualised. The solid cyan line depicts the median Balmer decrement value of our sample. In general, Balmer decrement values above five have significantly larger statistical uncertainties due to the low S/N nature of the spectra for which they are measured. The actual uncertainties are larger than visualised as the systematic uncertainties of the fit remain un-quantified. Sources with a measured Balmer decrement value above seven are likely to be non-physical and highlight the limitations of the software used in the low S/N spectra. We therefore recommend caution in the interpretation of the high Balmer decrement value tail in the histogram of Fig. \ref{fig:balmerdec}. The clustering of low uncertainty measurements around our sample median suggests that the overall sample measurement is robust.

The skewing of the H$\alpha$/H$\beta$ Balmer decrement to higher values in our sample is not unexpected as Type 2 AGN, by definition, have a large amount of dust present in their vicinity which obscures the central structure of the AGN. The large Balmer decrement values observed across the sample imply that dust in optically selected Type 2 AGN extends beyond the obscuring torus structure and into the NLR of the AGN. The presence of extended dust in the NLR may well be contributed to by polar dust components in the ionization cone of the AGN. These diffuse, elongated structures are interpreted as outflowing dusty winds, driven by radiation pressure \citep[][]{ramosalmeida2017}. In agreement with our result, \cite{lu2019} report an average NLR Balmer decrement of a sample of 554 optically selected broad line AGN to be 4.37$\pm$1.26 (dashed green line in Fig. \ref{fig:balmerdec}). \cite{jin2012} derive Balmer decrements in the range 1 - 12 with a mean value of 4.85$\pm$1.81 for narrow emission line components in a sample of X-ray selected Type 1 AGN and suggest that variations in electron density or dust abundance in the NLR is the cause of their higher than predicted value (dashed orange line in Fig. \ref{fig:balmerdec}). Similarly, \cite{purchnarewicz1997} find a mean Balmer decrement of 5$\pm$1 for a sample of 160 AGN from RIXOS, consisting of Type 1 and Type 2 X-ray selected AGN and attribute this result to dust present in their sources (dashed purple line in Fig. \ref{fig:balmerdec}). Some studies utilize the elevated Balmer decrement observed in Type 2 AGN as a means to classify AGN as Type 2 \citep[eg.][]{lehmann2001}.

\subsection{\oiii{} Line Luminosity Measurements}
\citet{reyes2008} catalogue of optically selected Type 2 quasars provides measurements of \oiii{} line luminosity. The measurements are provided with two separate methods; Gaussian, where the continuum over the wavelength range 4860 - 5060\AA{} is modelled with a linear fit and the \oiii{} emission line is modelled with a single Gaussian component, and semi-parametric where the \oiii{} line detected flux density is integrated and the linear continuum contribution is subtracted. \citet{reyes2008} find that their semi-parametric fits can often provide highly non-Gaussian solutions due to asymmetries in the line profile caused by radial outflows (eg. blue wings) and result in systematically higher line luminosities by $\sim$5$\%$. 

\begin{figure}
 \includegraphics[width=\columnwidth]{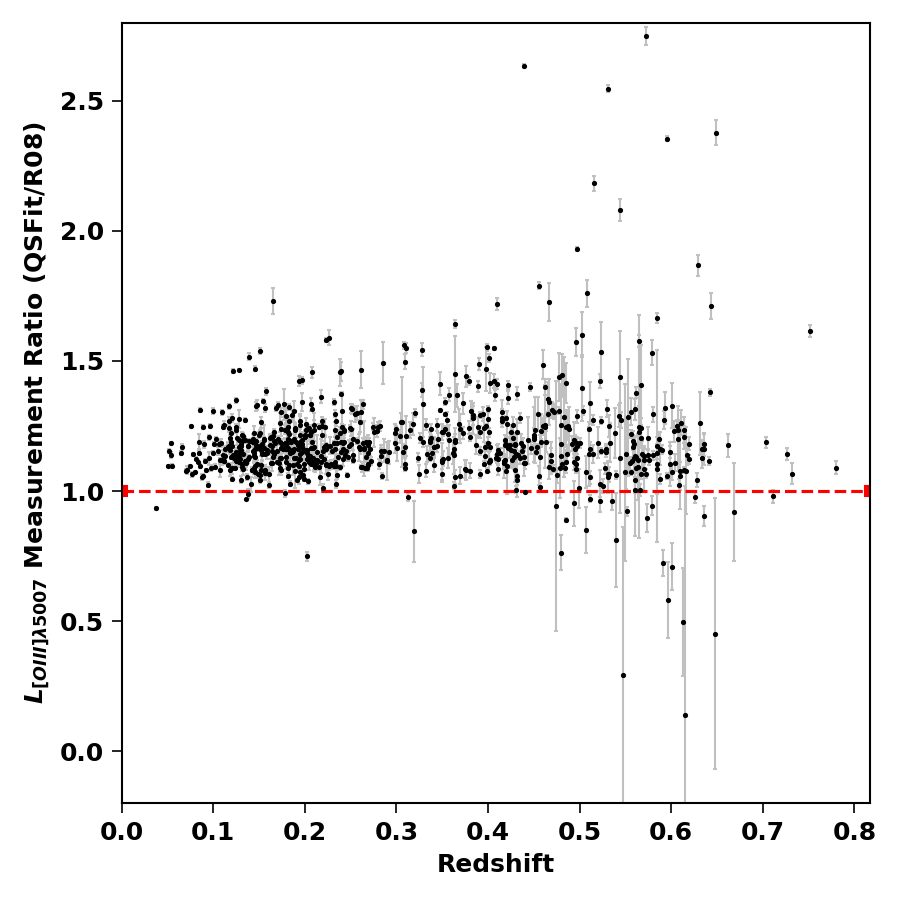}
 \caption{Ratio of \oiii{} emission line luminosity measurements made by \qsfit{} to those provided by \citet{reyes2008} semi-parametric fits for the same sample of Type 2 AGN vs redshift. The red dashed line shows the 1:1 ratio. A median systematic offset of 1.16 is explained by the differing emission line and continuum modelling employed by the two methods.}
 \label{fig:oiiiQ_R08}
\end{figure}

Fig. \ref{fig:oiiiQ_R08} depicts the ratio of \loiii{} as measured by \qsfit{} to that measured in \citet{reyes2008} with a semi-parametric fitting as a function of redshift. We find no significant trends or correlations among the quantities beyond the redshift induced correlation that arises when comparing pure measured values against each other. There exists a clear scatter between our \loiii{} measurements and those measured with a semi-parametric fit in \citet{reyes2008}. This scatter is expected from measurements made independently with two differing methods. The same scatter is observed when comparing our \loiii{} measurements to the Gaussian fitting method.

Alongside the intrinsic scatter between the measurements there is a systematic offset between \qsfit{} and \citet{reyes2008} semi-parametric measurements, with \qsfit{} providing larger \loiii{} values by a median ratio of 1.16. When compared to \citet{reyes2008}'s Gaussian \loiii{} measurements \qsfit{} again provides systematically larger measurements, with an increased median ratio of 1.22, consistent with the observation in \citet{reyes2008} that their semi-parametric method provides \loiii{} measurements that are systematically higher than their Gaussian method by $\sim$5$\%$. 

A factor that is likely to contribute to the systematic offset observed between \qsfit{} and \citet{reyes2008} \loiii{} measurements is the inclusion of an asymmetric blue wing component in our analysis of the line. The \oiii{} blue wing component in \qsfit{} is a second Gaussian component constrained to be placed at a shorter wavelength and have a larger FWHM than the main \oiii{} emission line to model the commonly observed asymmetric blue excess in the \oiii{} line, which is attributed to contributions to the line emission from radially outflowing gas components in the NLR (see section \ref{sec:knownemlines}). The blue wing component accounts for more than 10$\%$ of the total \oiii{} line flux in 707 of our spectra (88$\%$ of our sample). The addition of the blue wing component in our \oiii{} line modelling allows for a better fit to the profile of the line than a single Gaussian component can offer in most cases, ensuring that the line flux in the tails of the line profile are measured. 

To test if the better fit achieved with multiple \oiii{} components is generating the systematic offset between \qsfit{} and \citet{reyes2008} \loiii{} measurements, we repeat our \qsfit{} analysis of the sample with a modified procedure where the \oiii{} emission line is modelled with a single Gaussian component, mirroring the Gaussian method used in \citet{reyes2008}. When these constraints are applied, the systematic offset between \qsfit{} and \citet{reyes2008} \loiii{} measurements is reduced to a ratio of 1.03 when compared to \citet{reyes2008} semi-parametric method and 1.07 when compared to their Gaussian method. These results suggest that the majority of the systematic offset between \qsfit{} and \citet{reyes2008} \loiii{} measurements can be attributed to the inclusion of a blue wing component to the modelling of the \oiii{} emission line with \qsfit{}, which allows more line flux to be measured from the wings of the line. 

Despite accounting for the \oiii{} blue wing component, there is still a smaller systematic offset between the \loiii{} measurements of \qsfit{} and \citet{reyes2008}. In our analysis we model the entire continuum of each spectrum utilizing a single power-law for the AGN continuum contribution and a 5$\ $Gyr elliptical host-galaxy template \citep[from SWIRE template library;][]{polletta2007}. These components are then summed together to generate the full continuum. In contrast, \citet{reyes2008} model the continuum only in the wavelength range 4860 - 5060\AA{} with a linear fit. We confirmed that the remaining offset in measurements is associated with the differing continuum modelling methods used in \citet{reyes2008} and our work. 

\begin{figure*}
\includegraphics[height=0.27\paperheight,width=0.49\textwidth,keepaspectratio]{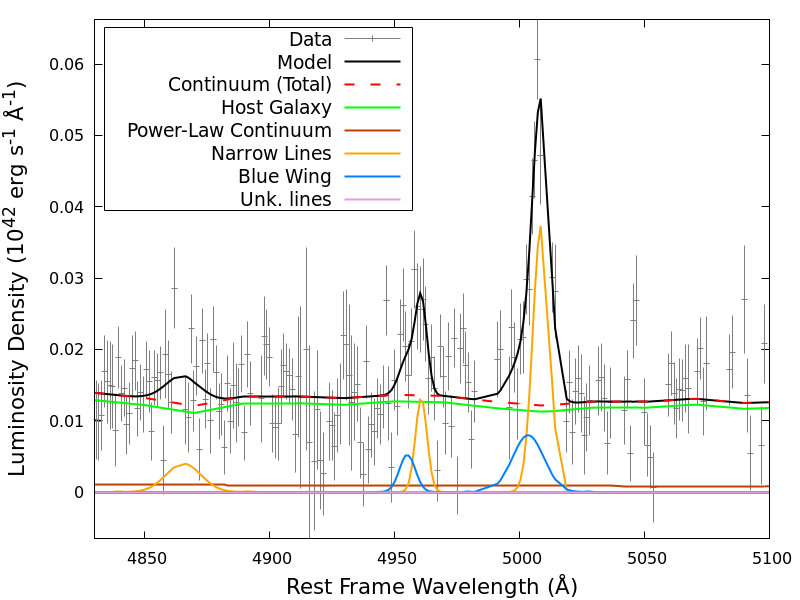}
\hfill
\includegraphics[height=0.27\paperheight,width=0.49\textwidth,keepaspectratio]{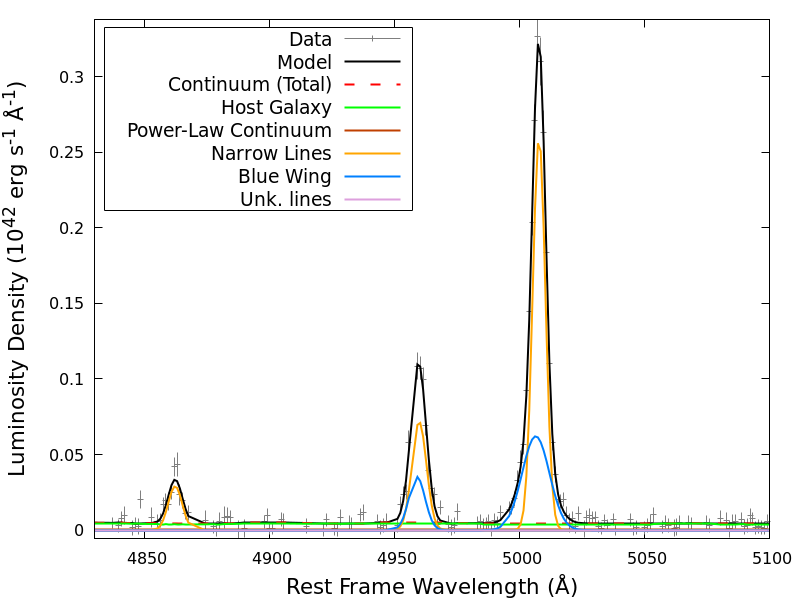}
\caption{Example of spectral fits of the [O {\sc iii}]+H$\beta$ complex in outliers in our comparison between \qsfit{} and \citet{reyes2008} (R08) measurements of \oiii{} emission line luminosity (see Fig~\ref{fig:oiiiQ_R08}), despite the fit being perfectly acceptable. \textit{Left}: SDSS J005813.75-001654.1, low S/N case (\qsfit{}/R08 ratio = 0.35). \textit{Right}: SDSS J151711.48+033100.1, high S/N case (\qsfit{}/R08 ratio = 0.49). Component colour scheme identical to Fig. \ref{fig:t2_recipe}.}
\label{fig:oiii_outlierexamples}
\end{figure*}

Outliers from the main locus of points in Fig \ref{fig:oiiiQ_R08} fall into 3 categories; (1) low S/N spectra where we still find our modelling acceptable for the data presented, (2) genuine acceptable fits to good quality data and (3) sources with complex \oiii{} profiles that require additional components to model. For the first and second cases the different fitting methods are responsible for the discrepancies. Fig. \ref{fig:oiii_outlierexamples} gives examples of a low S/N case (left) and a case where our fit is good despite the disagreement in values (right). We expect no significant differences in the spectra used between the analyses as the SDSS spectra were downloaded using the same PLATE-MJD-FIBER values. For the latter case, we present examples of these spectra in Fig. \ref{fig:strange_oiiicomplex}.

We derive the \oiii{} luminosity function (LF) following the 1/$V_{max}$ method \citep[][]{schmidt1968} employed in section 3 of \citet{reyes2008} using our \qsfit{} measurements of \loiii{}. We find that our derivation of the LF is compatible within errors to that derived using \citet{reyes2008} \loiii{} measurements. Therefore the systematic offset between the two measurement methods does not have an adverse affect on the calculated space density of luminous optical Type 2 AGN.

\section{SN Ratio}\label{sec:snratio}
\begin{figure}
 \includegraphics[width=\columnwidth]{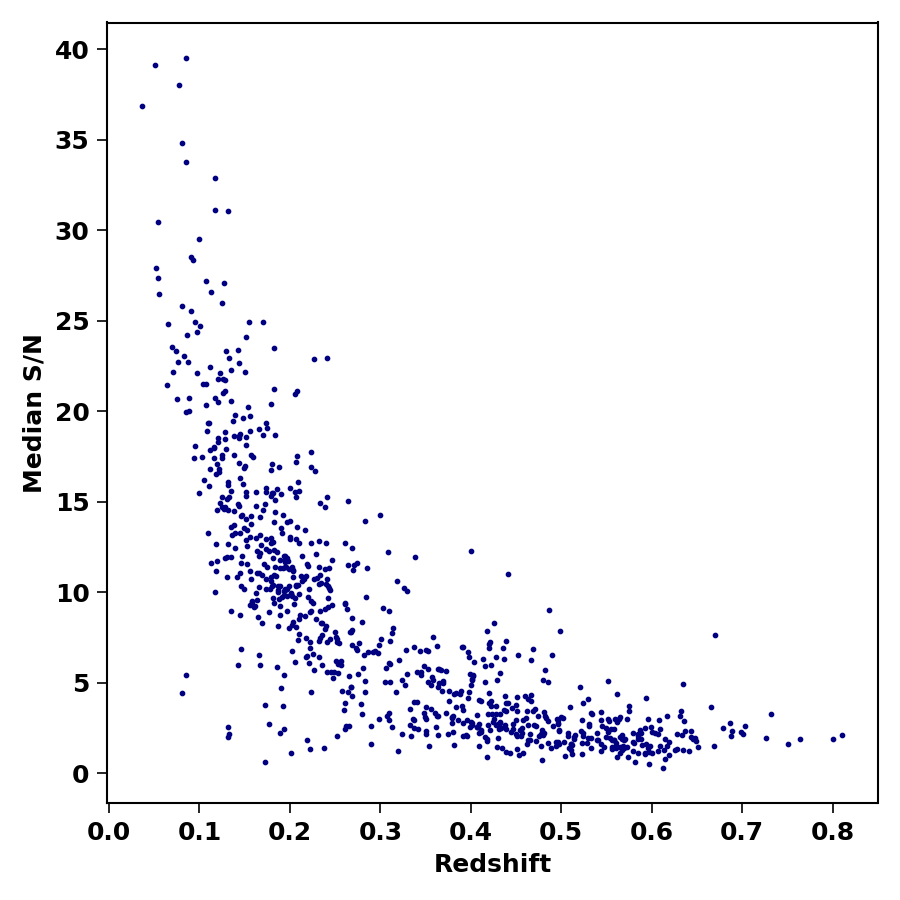}
 \caption{Median S/N value (median flux value / median uncertainty) of each spectrum in \citet{reyes2008} sample of Type 2 AGN as a function of source redshift.}
 \label{fig:snvsz}
\end{figure}

To investigate the effect of signal-to-noise ratio (S/N) on the measurements attained in our analysis with \qsfit{}, the S/N of each spectrum in the \citet{reyes2008} sample is estimated by dividing the median flux value of the spectrum by the median uncertainty. Using this S/N proxy it is found that the S/N of the SDSS spectra is a strong function of redshift, with the highest S/N spectra at low redshift and the S/N falling non-linearly as redshift increases as visualised in Fig. \ref{fig:snvsz}. The trend of falling S/N with redshift is an expected observation as all the spectra are generated with the same spectrograph and 3" aperture. Photon flux drops off approximately quadratically in the redshift range considered in this work, naturally resulting in the signal flux dropping off rapidly as redshift increases, while the detector noise remains approximately constant. 

\begin{figure}
 \includegraphics[width=\columnwidth]{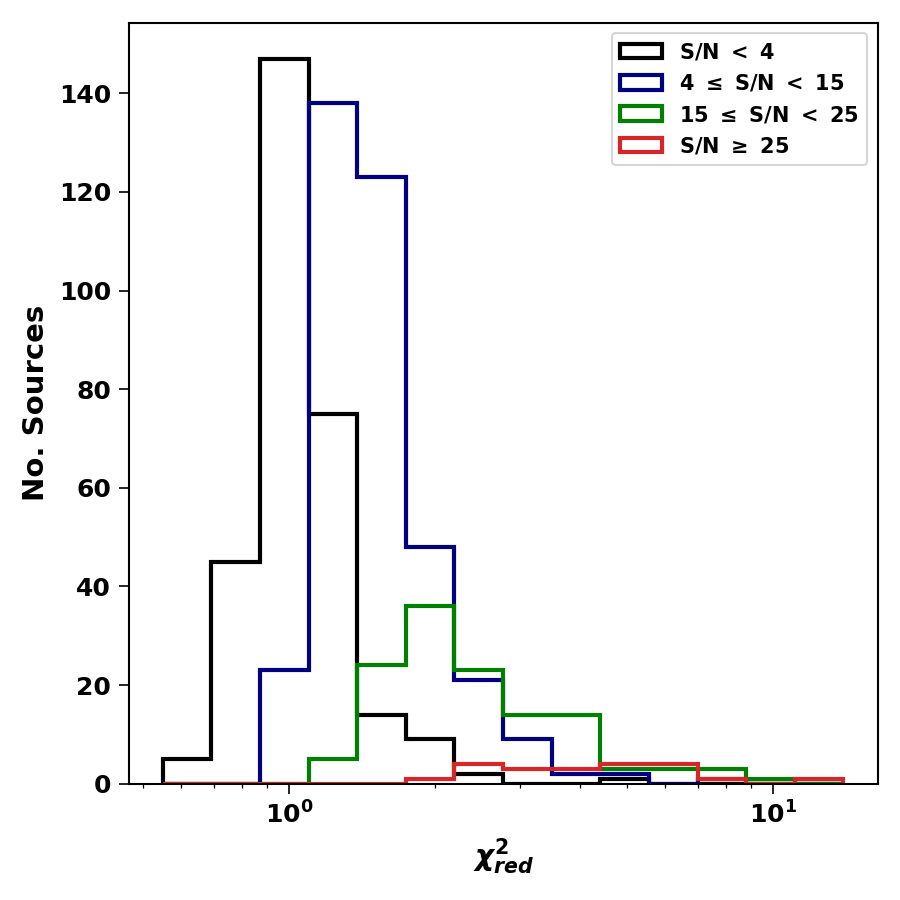}
 \caption{Distribution of reduced $\chi^{2}$ for global \qsfit{} spectral fits for our sample of 813 Type 2 AGN presented in \citet{reyes2008}, divided into signal-to-noise ratio (S/N) classes. \textit{Black}; sources with S/N $<$ 4. \textit{Blue}; sources with 4 $\geq$ S/N $<$ 15. \textit{Green}; sources with 15 $\geq$ S/N $<$ 25. \textit{Red}; sources with S/N $\geq$ 25.}
 \label{fig:snclasschi2}
\end{figure}

In order to estimate the impact of the S/N ratio on our analysis we divide our sample of luminous Type 2 AGN into 4 S/N groups; those with S/N $<$ 4 (298 sources),  4 $\leq$ S/N $<$ 15 (367 sources), 15 $\leq$ S/N $<$ 25 (126 sources) and those with S/N $\geq$ 25 (22 sources). Fig. \ref{fig:snclasschi2} presents a histogram of reduced $\chi^{2}$ fit statistics for global fits of the sample spectra in different S/N classes. We find that higher S/N spectra result in statistically worse fits, having a higher reduced $\chi^{2}$ than those with a lower S/N. This observation implies that spectra with a higher S/N require a more complex model to achieve a satisfactory fit, as weaker, often neglected emission lines are more prominent features in the data and strong emission line profiles are observed with increased fidelity, requiring multiple components to properly constrain the peaks of the lines. In higher S/N cases the above features become statistically more prominent since their spectra exhibit lower uncertainties, contributing to a higher reduced $\chi^{2}$ and highlighting the limitations of the model. On the contrary, for lower S/N spectra more prominent emission lines have generally lower fluxes as they originate from higher redshift sources and therefore have higher relative uncertainties, so are weighted less heavily in the calculation of the reduced $\chi^{2}$. 

\subsection{High S/N Recipe}
Motivated by statistically poor fits achieved with our \qsfit{} Type 2 AGN recipe for the highest S/N spectra, we develop and present a specialised recipe for high S/N (HSNR) spectra. 

Reduced $\chi^{2}$ is a statistic intended to assess the fit of a global model to the data it attempts to describe. Therefore, to assess the performance of a model in specific portions of our spectral fits we adopt the proxy metric $\chi^{2}_{portion}/N_{S}$, where $\chi^{2}_{portion}$ is the total $\chi^{2}$ statistic of the samples in the portion of the fit being analysed and $N_{S}$ represents the number of spectral channels covered by the examined portion.

Via analysis of $\chi^{2}_{portion}/N_{S}$ of different components of our spectral fits we find that in all cases, particularly so for HSNR spectra, the most prominent lines in the spectrum are statistically the worst fit components, suggesting that in the HSNR case a single Gaussian component is insufficient to constrain the profile of prominent emission lines. The focus of this specialised HSNR recipe, then, is to more flexibly model the most prominent lines encountered in Type 2 AGN spectra.

The [O {\sc iii}]$\lambda$4959, \oiii{} and H$\alpha$ emission lines are identified as the most frequently observed prominent lines in our sample, which prove to be insufficiently fit with a single Gaussian component in HSNR spectra. To create a more flexible line profile we try modifying our Type 2 AGN recipe to fit these most prominent lines using; (1) two Gaussian components, (2) a single Lorentzian component, (3) two Lorentzian components and, for completeness, (4) a single Voigt component. In all cases the number of unknown lines considered in the fit is increased from the two used in our standard Type 2 recipe to six. This modification is made because for HSNR spectra the increased protrusion of weaker emission lines above the noise level means that unknown lines are more likely to be used as intended, at the site of an emission line not considered in our recipe, rather than being used to fill broad residuals.

\begin{figure}
 \includegraphics[width=\columnwidth]{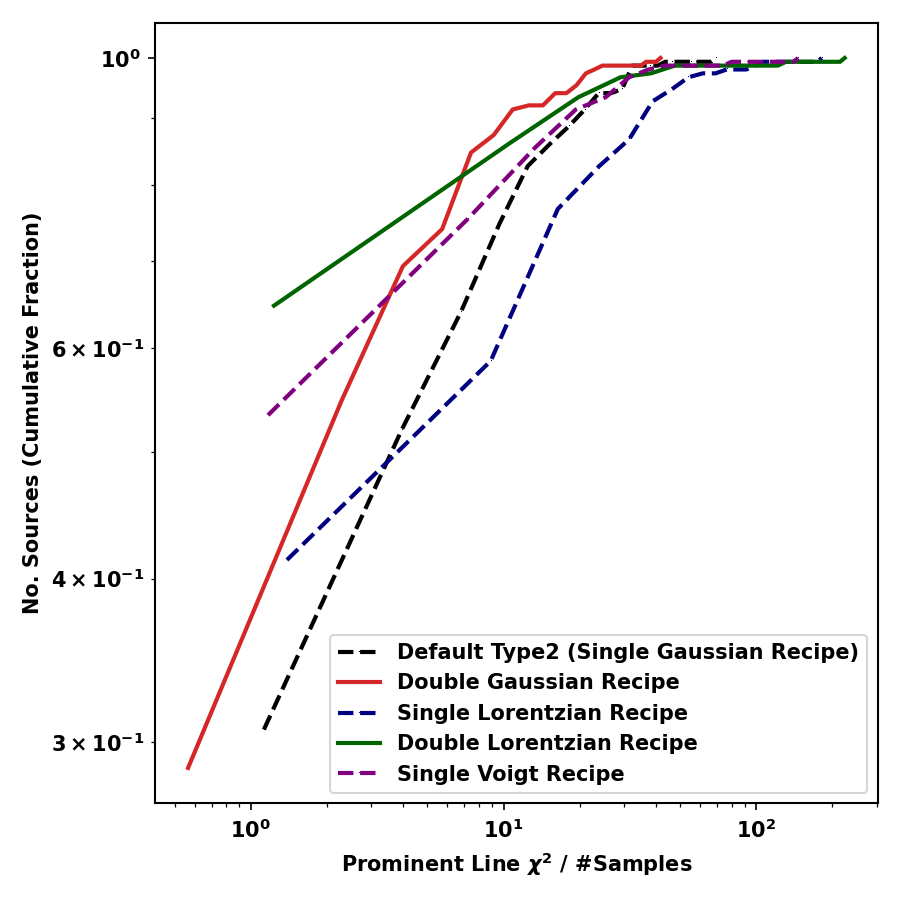}
 \caption{Cumulative histogram of $\chi^{2}$ divided by the number of samples for \qsfit{} spectral fitting results around the H$\alpha$ and H$\beta$ emission line complexes for five emission line profile variations for sources with median S/N $\geq$ 15. Emission line profiles of [O {\sc iii}]$\lambda$4959, \oiii{} and H$\alpha$ are varied in each case to be modelled by; (black) single Gaussian component, (red) double Gaussian, (blue) single Lorentzian component, (green) double Lorentzian, (purple) single Voigt component. }
 \label{fig:hsnrSLchi2}
\end{figure}

We fit the spectra in our sample with S/N $\geq$ 15 (148 sources) with each variation of our prominent line recipe modifications discussed above. Fig. \ref{fig:hsnrSLchi2} presents the cumulative $\chi^{2}_{portion}/N_{S}$ distributions around the H$\beta$ and H$\alpha$ complexes for each of our recipe modifications. In each case the $\chi^{2}_{portion}/N_{S}$ values presented are calculated for $\pm$ 3 FWHM around the center of the emission lines [O {\sc iii}]$\lambda$4959, \oiii{} and H$\alpha$ in each spectrum. 

Fig. \ref{fig:hsnrSLchi2} shows that fitting [O {\sc iii}]$\lambda$4959, \oiii{} and H$\alpha$ with a single Lorentzian line profile provides statistically worse fits than fitting these lines with a single Gaussian component in HSNR cases, mirroring our findings for all cases (see Sect. \ref{sec:gvslprofiles}). We also see that modelling [O {\sc iii}]$\lambda$4959, \oiii{} and H$\alpha$ with two Gaussian components provides a significantly improved fit in the most prominent lines of HSNR spectra, with two Lorentzian components also showing an improved fit compared to a single Gaussian component, yet not as successful over the whole sample as modelling with two Gaussians. Fitting prominent lines with a single Voigt profile is found to give marginally worse performance compared to the fits using two Lorentzian components. Table \ref{tab:hsnr_variations} provides the $\chi^{2}_{portion}/N_{S}$ statistic values for global model fits and prominent line fits for the 148 spectra with S/N $\geq$ 15 analysed. The values in Table \ref{tab:hsnr_variations} quantitatively back up our findings discussed above, showing that modelling [O {\sc iii}]$\lambda$4959, \oiii{} and H$\alpha$ with two Gaussian components provides the statistically best fits globally and around the most prominent line complexes of all the modelling variations considered in this section. These results highlight the necessity to model the most prominent emission lines in HSNR spectra with multiple components. 

\begin{table*}
 \caption{Performance of different line profile combinations to model [O {\sc iii}]$\lambda$4959, \oiii{} and H$\alpha$ emission lines in high S/N sources (median S/N $\geq$ 15). We use $\chi^{2}$ or the total $\chi^{2}$ of a portion of the fit ($\chi^{2}_{portion}$) divided by the number of samples the considered range covers ($N_{S}$) as a proxy metric for the reduced $\chi^{2}$. Prominent line $\chi^{2}_{portion}/N_{S}$ refers to our proxy metric as calculated for a portion of the fit $\pm$3 line FWHM around the emission line centroids of [O {\sc iii}]$\lambda$4959, \oiii{} and H$\alpha$ in each spectrum.}
 \label{tab:hsnr_variations}
 \centering
 \begin{tabular}{lccc}
  \hline
  Recipe & Median Global Fit $\chi^{2}/N_{S}$ & Median Prominent Line $\chi^{2}_{portion}/N_{S}$  \\
  \hline
  Default Type 2 & 2.55$\pm$0.82 & 6.46$\pm$3.76 \\
  Double Gaussian & 2.17$\pm$0.58 & 3.60$\pm$1.89 \\
  Single Lorentzian & 2.95$\pm$1.04 & 12.90$\pm$7.96 \\
  Double Lorentzian & 3.03$\pm$1.05 & 7.43$\pm$3.99 \\
  Single Voigt & 2.46$\pm$0.78 & 6.87$\pm$3.86 \\
  \hline
 \end{tabular}
\end{table*}

\section{Double Peaked Sources}\label{sec:dp}
AGN spectra hosting double peaked narrow emission lines are rare, with $\sim$1$\%$ of low-z Type 2 AGN spectra exhibiting such features in optically selected samples \citep[][]{shen2011}. The double-peaked emission line features can be generated by kinematically disturbed NLRs induced by bi-conical outflows and disk-like rotation of the NLR or alternatively by a merging pair of AGNs where there is relative motion between the two NLRs. The latter of these scenarios is of central interest in the context of studying galaxy mergers and AGN fuelling, however the lack of unambigiously detected objects of this kind ($\leq$0.1$\%$ of SDSS quasars) is in tension with observed and simulation-based galaxy merger rates. Literature in the field is conflicted on the favoured origin for the double peaked narrow emission line phenomenon in AGN; some works favour disturbed NLR situations \citep[eg.][]{wang2019, shen2011} while other authors favour AGN binaries \citep[eg.][]{wang2009} and some works stress ambiguity with no clear evidence for either scenario \citep[eg.][]{smith2010, liu2010}.

Throughout the testing of our Type 2 AGN fitting procedure a significant portion of the outliers in the self-consistency checks presented in Sect. \ref{sec:sample_results} have been from spectra that result in a poor fit due to the presence of double peaked narrow emission lines. These features are not accounted for with our default \qsfit{} recipe. Motivated by the interesting physics found in such sources and the lack of a publicly available, generalised fitting routine for double peaked sources we present a specialised \qsfit{} recipe for fitting Type 2 AGN spectra that host double peaked emission lines.

\cite{liu2010} presented a sample of 167 Type 2 AGN with double peaked [O {\sc iii}]$\lambda$4959 and \oiii{} lines selected from SDSS DR7. The sample is derived from a parent MPA-JHU SDSS DR7 emission line galaxy sample with z $<$ 0.7, selecting candidates based on [O {\sc iii}] emission line equivalent width, BPT emission line ratio cuts and a S/N constraint of median S/N $>$ 5~pixel$^{-1}$. The emission line galaxy sample is then supplemented with $\sim$400 Type 2 AGN from the \citet{reyes2008} sample, creating a total parent sample of 14,756 spectra. Each spectrum has then been visually inspected to identify the 167 objects with unambiguously detected double peaked [O {\sc iii}] lines, requiring two components to fit. 

We matched the \cite{liu2010} double peaked source catalogue of 167 double peaked emission line objects to the \citet{reyes2008} Type 2 AGN sample analysed in this work. We found an intersection of 51 objects. This intersection constitutes 6.7$\%$ of our Type 2 AGN sample, which is a significantly higher fraction than expected from the prevalence of double peaked sources in the overall SDSS. This discrepancy may suggest that the double peaked emission line phenomenon is more common in the high luminosity Type 2 AGN ($L_{bol} \gtrsim$ 10$^{44}$~ergs$^{-1}$) examined in this work. In the case of the bi-conical outflow origin of double peaked emission lines this is consistent as higher luminosity AGN distribute more energy into their surroundings, producing more powerful radiatively driven outflows at a greater rate than their lower luminosity counterparts. Six of these spectra have a significant fraction ($>50\%$) of their spectral channel flagged as possibly affected by issues (see Section \ref{sec:sample}), hence we neglected them. We therefore have 45 double peaked Type 2 AGN spectra which are adequate for fitting with \qsfit{}. This sample is henceforth used as our sample of double peaked Type 2 AGN sources.

\subsection{Double Peaked Type 2 AGN Recipe}

Our double peaked Type 2 AGN recipe aims to provide a general routine to produce acceptable fits to the majority of sources in our sample, allowing the characteristics of the sample to be quickly examined.

We model the [O {\sc ii}]$\lambda$3727 doublet, H$\beta$, [O {\sc iii}]$\lambda$4959, \oiii{}, H$\alpha$ and [N {\sc ii}]$\lambda$6583 emission lines using two narrow line Gaussian components. These are the highest luminosity lines apparent in our sample of double peaked Type 2 AGN optical spectra for which the red and blue components can be de-blended. Narrow lines used to model red and blue peaks in double peaked emission lines have FWHM constrained to the range [100 - 500]$\ $km~s$^{-1}$. This constraint ensures that the components model the peaks of the line profile as intended, rather than one Gaussian dominating the whole structure and the second being used to account for asymmetries. In each case the second narrow component which is newly added to the \qsfit{} recipe has velocity offset constrained to be greater than its counterpart, placing it on the shorter wavelength (blue) side. 

An additional `core' narrow line component is utilised in the modelling of [O {\sc iii}]$\lambda$4959 and \oiii{} double peaked emission lines. The core component is constrained to have FWHM in the range [100 - 1000]$\ $kms$^{-1}$. This component accounts for asymmetries in the base of the luminous [O {\sc iii}] lines, allowing the red and blue narrow line components to model the double peaked features as intended, rather than filling the base of the line which contributes more strongly to the overall $\chi^{2}$ statistic of the fit. 

During fitting, we patch the velocity offset and FWHM of each separate component (red peak, blue peak and core, for [O {\sc iii}] lines) of [O {\sc ii}]$\lambda$3727, [O {\sc iii}]$\lambda$4959 and [N {\sc ii}]$\lambda$6583 to those of \oiii{}. The velocity offset and FWHM parameters of H$\alpha$ are also patched to those of H$\beta$. From a physical standpoint, the red, blue and core components of each double peaked forbidden or permitted transition is emitted from the same gas under identical dynamical conditions, meaning that a simultaneous fit is achievable for each of these parameters. Applying the same velocity offsets and FWHM to red and blue double peaked components is particularly important for H$\alpha$ and [N {\sc ii}]$\lambda$6583, which are often blended together in the H$\alpha$ complex for double peaked sources. Patching their velocity offsets from the systematic velocity and FWHM to lines outside of the complex allows their decomposition to be estimated where they would otherwise become degenerate parameters.

\begin{figure*}
\includegraphics[height=0.27\paperheight,width=0.49\textwidth,keepaspectratio]{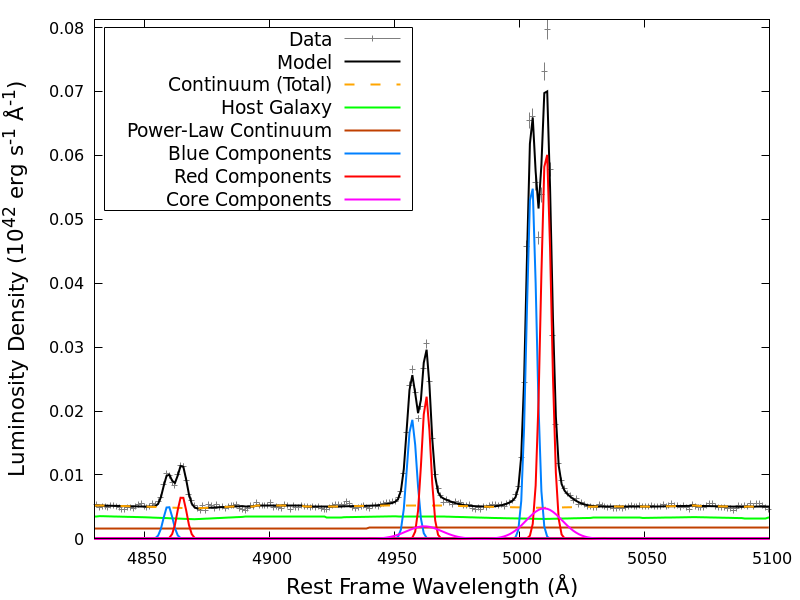}
\hfill
\includegraphics[height=0.27\paperheight,width=0.49\textwidth,keepaspectratio]{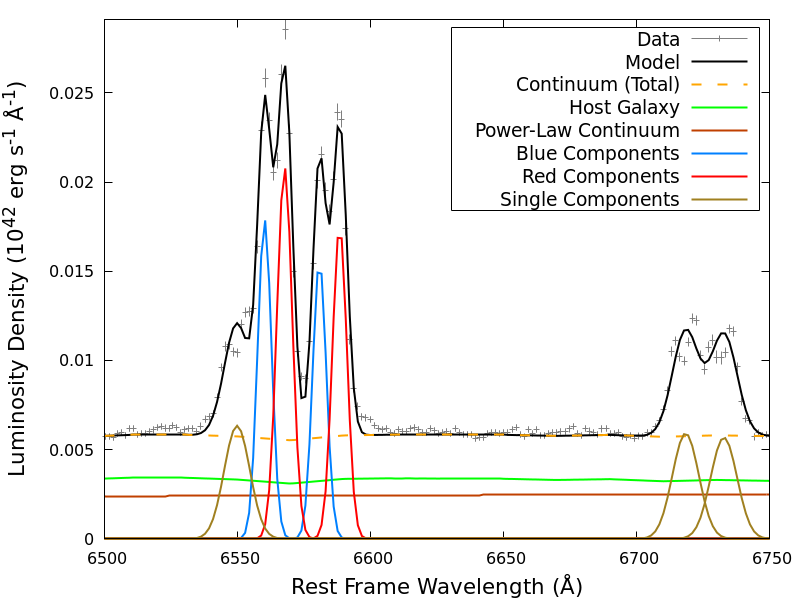}
\vfill
\caption{Example spectral decomposition of the [O {\sc iii}]+H$\beta$ (left) and H$\alpha$+[N {\sc ii}] (right) complexes in a double peaked Type 2 AGN SDSS spectrum (SDSS J080218.64+304622.7) using our specialised double peaked Type 2 AGN \qsfit{} fitting procedure. The velocity separation of the red and blue peaks in \oiii{} is 346~kms$^{-1}$. The solid black line represents the overall fit model, the solid red and blue lines show the red and blue components of double peaked lines, while the magenta lines depict the core components of these lines. The host-galaxy template is shown in solid green and the power-law AGN continuum in solid dark orange. The dashed orange line presents the overall continuum level given by the sum of the host-galaxy and AGN continuum, while the solid orange lines represent the single component known narrow emission lines considered in this fitting recipe.}
\label{fig:dp_decomposition}
\end{figure*}

Fig. \ref{fig:dp_decomposition} gives an example spectral fit of the H$\beta$ and H$\alpha$ complexes of a double peaked source using our \qsfit double peaked Type 2 AGN recipe.

\subsection{Double Peaked Results}

\begin{figure}
 \includegraphics[width=\columnwidth]{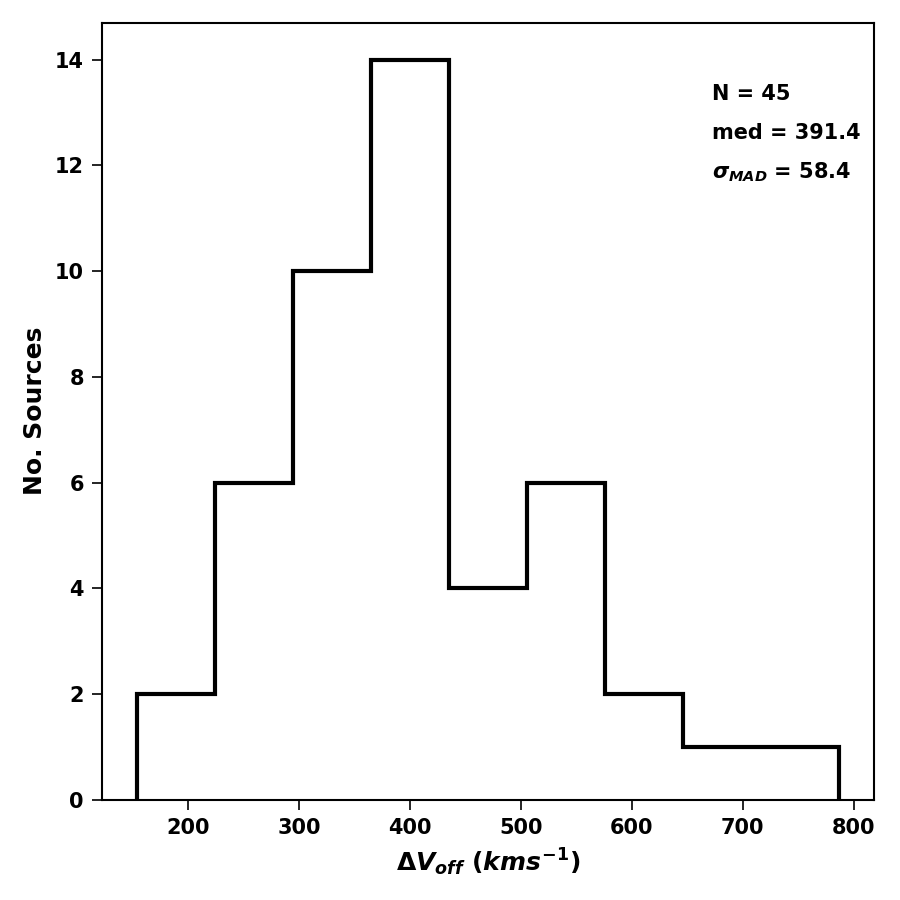}
 \caption{Histogram showing the distribution of separation velocities between the red and blue peaks of the \oiii{} line in double peaked Type 2 AGN spectra analysed with \qsfit{} in this work.}
 \label{fig:dp_sep}
\end{figure}

\begin{figure}
\includegraphics[width=0.48\textwidth,keepaspectratio]{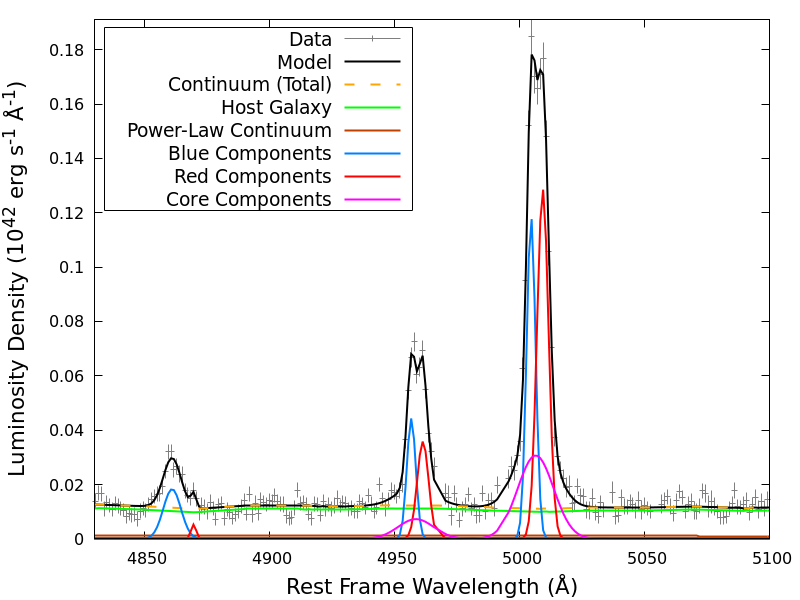}
\vfill
\includegraphics[width=0.48\textwidth,keepaspectratio]{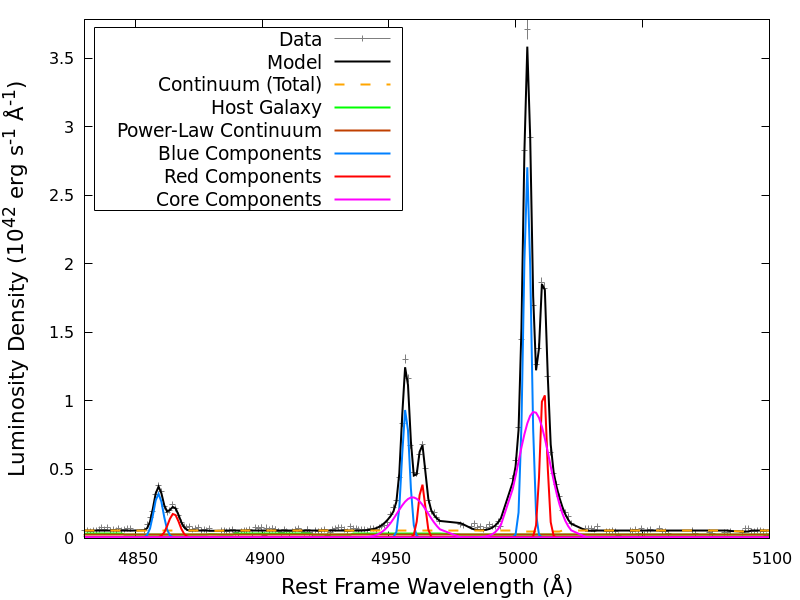}
\vfill
\includegraphics[width=0.48\textwidth,keepaspectratio]{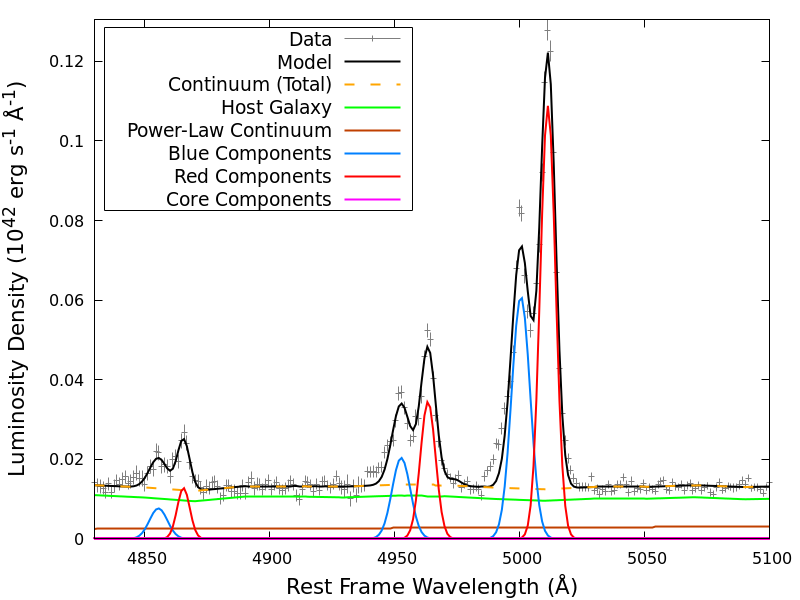}
\caption{Three example \qsfit{} spectral decompositions of double peaked Type 2 AGN SDSS spectra with varying red and blue component separations ($\Delta$v). (Top) SDSS J095005.82+481338.5, $\Delta$v = 278$\ $kms$^{-1}$. (Center) SDSS J153944.12+343503.9, $\Delta$v = 364$\ $kms$^{-1}$. (Bottom) SDSS J122709.83+124854.5, $\Delta$v = 623$\ $kms$^{-1}$. Component colour scheme identical to Fig. \ref{fig:dp_decomposition}. }
\label{fig:dp_sep_examples}
\end{figure}

Fig. \ref{fig:dp_sep} presents a histogram of the separations between red and blue peaks of \oiii{} in our sample of double peaked Type 2 AGN spectra. We find a median separation of 390$\pm$60$\ $kms$^{-1}$. The shape of the distribution shows that sources with a peak separation larger than the median become progressively more rare to observe as the separation becomes larger. This large value decline could be due to the scarcity of bi-conical outflows with high enough velocities to produce this separation observed at orientation angles for which the double-peaked lines are apparent in the data. We also note a sharp decline in the number of sources observed as the separation decreases from the median. The small separation decline is likely explained by the visual sample selection method employed in \citet{liu2010}, which is biased against smaller peak separations due to the lines become blended.  Indeed, no separation is observed for values lower than 150$\ $kms$^{-1}$ which is the instrumental resolution FWHM of SDSS spectra; double peaked lines cannot possibly be de-blended with separations lower than this value. Fig. \ref{fig:dp_sep_examples} shows examples of the [O {\sc iii}] complex fits of double peaked spectra with small (278$\ $kms$^{-1}$), median (364$\ $kms$^{-1}$) and large (623$\ $kms$^{-1}$) \oiii{} peak separations. 

\begin{figure}
 \includegraphics[width=\columnwidth]{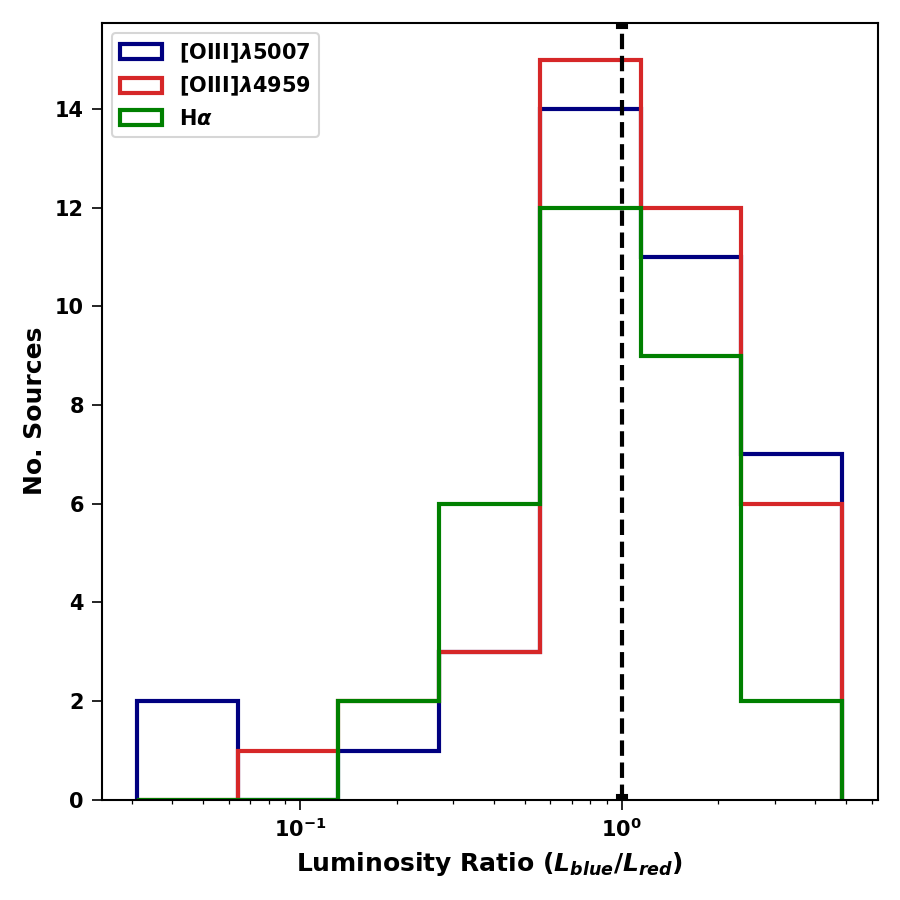}
 \caption{Histogram showing the distribution of luminosity ratios (L$_{blue}$/L$_{red}$) between the blue and red peaks of the [O {\sc iii}]$\lambda$4959, \oiii{} and H$\alpha$ emission lines (shown in red, blue and green, respectively) in double peaked Type 2 AGN spectra analysed with \qsfit{} in this work. The black dashed line shows the 1:1 ratio for which the red and blue peaks have equal luminosity.}
 \label{fig:dp_luminosityratio}
\end{figure}

Fig. \ref{fig:dp_luminosityratio} depicts the luminosity ratio between blue and red components (L$_{blue}$/L$_{red}$) of double peaked [O {\sc iii}]$\lambda$4959 (red), \oiii{} (blue) and H$\alpha$ (green) emission lines. Median luminosity ratios of 1.1$\pm$0.6, 1.0$\pm$0.5 and 0.8$\pm$0.4 are observed for [O {\sc iii}]$\lambda$4959, \oiii{} and H$\alpha$ respectively. All values are consistent with unity within uncertainty across the sample, showing that there is no preference for the red or blue peaks to be systematically more luminous in double peaked Type 2 AGN. The similarity in distribution across all three lines suggests that the luminosity ratio for forbidden and permitted lines are consistent in double peaked sources. 

\subsubsection{Emission Line Origin}

The measured velocity offsets of 200 - 600 kms$^{-1}$ are compatible with both galactic winds driven by AGN feedback and winds driven by stellar processes \citep[see][for a review]{veilleux2005}. Our median value of 390$\pm$60$\ $kms$^{-1}$ is a relatively low velocity for an AGN driven outflow. This is expected from the fact we are examining Type 2 AGN which are generally less luminous than their Type 1 counterparts. Indeed, using the \loiii{} bolometric correction of \citet{lamastra2009} we find a median bolometric luminosity of our main sample to be $L_{bol}\sim$10$^{45}$~ergs$^{-1}$. Only 2.6$\%$ of our sources have $L_{bol} \geq$ 10$^{46}$~ergs$^{-1}$, holding the potential to drive the strongest outflows.

\begin{figure}
 \includegraphics[width=\columnwidth]{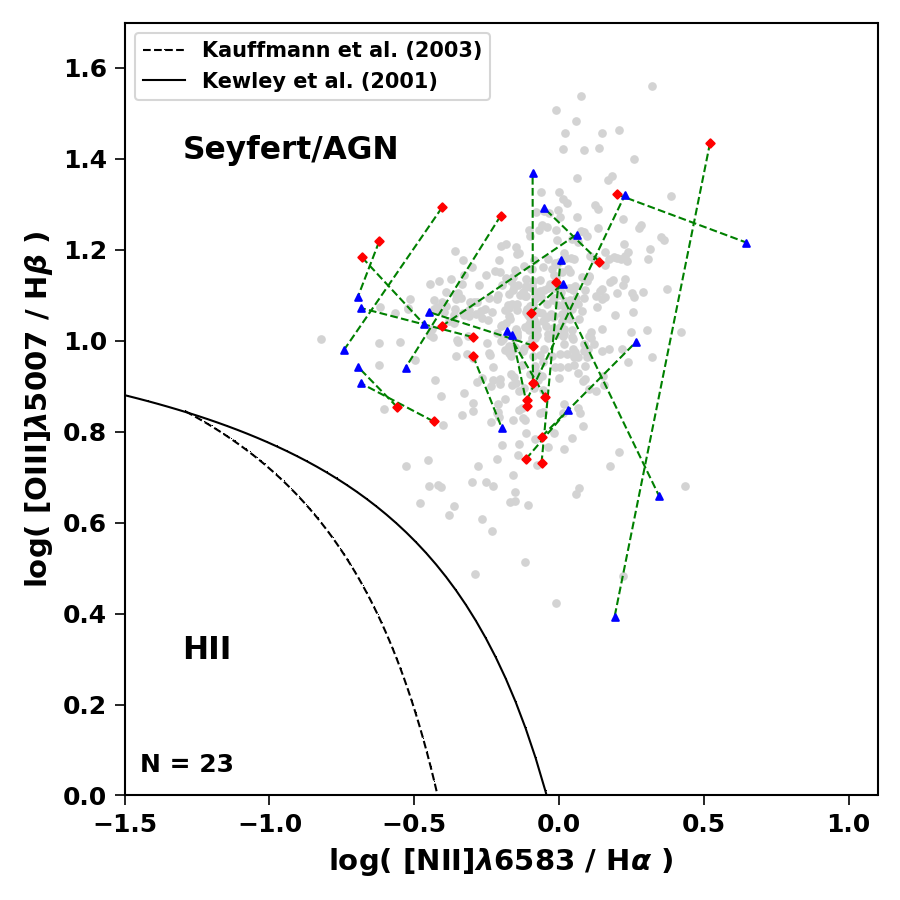}
 \caption{\oiii{}/H$\beta$ vs [N {\sc ii}]$\lambda$6583/H$\alpha$ BPT diagram. Double peaked Type 2 AGN red components are plotted with red diamonds and blue components are represented by blue triangles for the 23 sources for which \qsfit{} measurements of all 8 lines are available. Green dashed lines connect the red and blue component positions for each individual Type 2 AGN spectrum. The BPT positions of our main sample of Type 2 AGN spectra are plotted in grey. The solid curve depicts the theoretical maximum starburst line dividing AGN from HII regions (star-forming galaxies) derived by \citet{kewley2001}, the dashed curve defines the empirical AGN-HII region demarcation introduced by \citet{kauffmann2003}.}
 \label{fig:dp_bpt}
\end{figure}

We plot the red and blue double peaked components for which the \oiii{} and H$\alpha$ complexes are present in the spectrum on the \oiii{}/H$\beta$ vs [N {\sc ii}]$\lambda$6583/H$\alpha$ BPT diagram. This is shown in Fig. \ref{fig:dp_bpt} for which the blue component positions are given as blue triangles and the red component positions are given as red diamonds. The green dashed lines link the blue and red component positions for individual double peaked sources and the grey points show the positions of all the single-peak treated sources in our \citet{reyes2008} sample. The positions of red and blue double peaked components are consistent with the overall single peaked sample positions and none of the sources change classification. The consistency shows that both the red and blue peaks of all our double peaked sources are ionized by AGN continuum emission. There appears to be no preference for red or blue components to have systematically higher or lower \oiii{}/H$\beta$ or [N {\sc ii}]$\lambda$6583/H$\alpha$ ratios. The lack of trend suggests that there is no preference for red or blue components to have higher ionization parameter, as traced to first order by \oiii{}/H$\beta$, or metallicity as traced to first order by [N {\sc ii}]$\lambda$6583/H$\alpha$ in the Seyfert/LINER branch \citep[][]{agostino2021}.

From the properties explored in this section we find our double peaked sample results to be consistent with either bi-conical outflow or dual AGN origins.  

\subsubsection{Host-Galaxy Morphology}
\begin{figure*}
\includegraphics[width=0.2\textwidth,keepaspectratio]{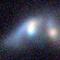}
\hfill
\includegraphics[width=0.2\textwidth,keepaspectratio]{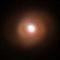}
\hfill
\includegraphics[width=0.2\textwidth,keepaspectratio]{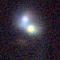}
\hfill
\includegraphics[width=0.2\textwidth,keepaspectratio]{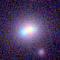}
\caption{Four example 15" Pan-STARRS1 images used to assess host-galaxy morphology for merger signatures. Each example exhibits one of the features visually inspected for in our analysis. Merger signature presented (from left to right); tidal features, ring, close companion, colour gradient.}
\label{fig:dp_cutout_examples}
\end{figure*}

To provide a basic insight into the origins of double peaked emission lines in our sample we visually inspect 15" $grz$ Pan-STARRS1 \citep[][]{chambers2016} cutout morphology for the prevalence of galaxy merger evidence in our sample. Alongside cutouts of our sample of 45 successfully fit double peaked Type 2 AGN, we create three additional samples comprised of 45 cutouts of single peaked Type 2 AGN from our main sample. Each of these additional samples are matched in \oiii{} luminosity and redshift with our double peaked sample and contain a unique selection of sources. We perform identical inspection analyses on all four samples, inspecting for several morphological features related to major and minor galaxy mergers; tidal features, rings, close companions at the redshift of the AGN host and multiple distinct stellar populations within the galaxy traced by colour gradients. Fig. \ref{fig:dp_cutout_examples} gives examples of positive results for each of these inspection categories.

Table \ref{tab:morphology} presents the results of this exercise. We find in all inspection categories our double peaked sample presents merger feature rates consistent with those derived from the parent sample of single peaked sources. In the case of close companions in the frame of the 15" image our double peaked sample has the least occurrences with 16\% (7 sources) displaying such a feature compared to an average of 27\% among the three single peaked samples. To the resolution available (64$\ $kpc at the median redshift of our sample, $z$=0.28), a minority of double peaked AGN host galaxies show evidence of merger activity. This suggests that the phenomenon is not directly related to galaxy mergers and is likely largely caused by bi-conical outflows or NLR kinematics, rather than dual AGN. This is not to say that there are no dual AGN present in the sample, however higher resolution and multi-wavelength data would be needed to draw conclusions on the nature of individual objects, which is beyond the scope of this paper. Thus, to the resolution available in Pan-STARRS1 imaging our double peaked Type 2 AGN are found to reside in galaxies with similar rates of morphological merger signatures as their single peaked counterparts. This suggests that the majority of double peaked sources in our sample have a bi-conical outflow origin rather than a dual AGN origin.

\begin{table*}
 \caption{Visual inspection of 15" Pan-STARRS1 images for four categories of merger activity. Inspection is completed for our fitted sample of 45 double peaked Type 2 AGN and three unique samples of 45 single peaked Type 2 AGN from our parent sample matched in \oiii{} luminosity and redshift. Absolute counts are given the table with the percentage of the inspection sample this constitutes to given in brackets for each value.}
 \label{tab:morphology}
 \centering
 \begin{tabular}{lccccc}
  \hline
  Sample (each with 45 sources) & Tidal & Ring & Companion & Colour Gradient  \\
  \hline
  Double Peaked & 7 (16\%) & 2 (4\%) & 7 (16\%) & 9 (20\%)\\
  Single Peaked 1 & 6 (13\%)& 5 (11\%) & 14 (31\%) & 11 (24\%)\\
  Single Peaked 2 & 5 (11\%) & 3 (7\%) & 12 (27\%)& 7 (16\%)\\
  Single Peaked 3 & 5 (11\%) & 4 (9\%) & 11 (24\%) & 7 (16\%)\\
  \hline
 \end{tabular}
\end{table*}

A similar analysis was performed in \citet{comerford2013} where visual inspection of morphology in SDSS $gri$ imaging of dual AGN candidates that exhibit double peaks or emission line offsets from the system redshift selected from the AGES survey \citep[][]{kochanek2012}. They found that dual AGN candidates are found to have close (within 5") companions in 29\% of sources compared to 9\% in their parent sample of Type 2 AGN. This appears to be at odds with our analysis, however the frequency of merger features in their dual AGN candidates may be affected by their small sample size of seven sources. We use a sample six times larger in our analysis and therefore argue that our results give a more complete picture of the merger rates in double peaked AGN. A 29\% fraction of mergers is consistent with the value we find for our single-peaked AGN inspection sub-samples. As our parent sample is selected to consist of high luminosity Type 2 AGN this could suggest that high luminosity AGN are found more frequently in galaxies with merger signatures when compared with the overall Type 2 AGN population.

\section{Summary}
Motivated by the need to rapidly process and classify large volumes of AGN spectra from upcoming astronomical facilities we developed an optical Type 2 AGN spectral fitting routine optimised for large volume analysis using the scalable and automatic spectral fitting software \qsfit{}. The development of this recipe allows Type 2 AGN spectra to be analysed in a consistent fashion to Type 1 AGN spectra within the same software package, eliminating systematic biases when comparing the values derived from the fits. To qualify its performance and carry out some initial scientific studies our presented fitting routine is applied to a sample of 813 optically selected luminous Type 2 AGN spectra from the SDSS drawn from the sample collated in \citet{reyes2008}. The median time taken to analyse a low-z SDSS Type 2 AGN spectrum with our code is 1.6$\pm$0.9$\ $s. 

We find no significant statistical preference to use Gaussian or Lorentzian emission line profiles for low-z Type 2 AGN SDSS spectra. Gaussian and Lorentzian profiles give median sample reduced $\chi^{2}$ values of 1.39$\pm$0.36 and 1.45$\pm$0.38, respectively, which are consistent with one another within uncertainty. 

To confirm the reliability of the measurements made using our fitting routine we plot the measured positions of our sources on the BPT emission line diagnostic diagrams. We observe that the bulk of our measurements lie in the expected AGN/Seyfert portion of these diagrams confirming that the ionizing continuum generating the emission lines are of AGN origin. Outliers from the main locus of sources are well explained by spectra exhibiting low S/N ratios or double peaked emission lines, for which our default Type 2 AGN model is incomplete. Alongside this test we leave the known 3:1 ratio of \oiii{} to [O{\sc iii}]$\lambda$4959 emission line flux as a free parameter in our fit (for the core components of the lines) to check if the correct value is recovered. We find a median [O{\sc iii}] ratio of 3.1$\pm$0.1 across our sample, compatible with expectations and the literature.

The Balmer decrement (H$\alpha$/H$\beta$ emission line flux ratio) is calculated for a subset of 456 sources in our sample for which both associated Balmer emission lines are available. We find a median value of 4.5$\pm$0.8 across our sample of luminous Type 2 AGN. This value is consistent with other works that calculate the NLR Balmer decrement for samples of Type 1 and Type 2 AGN, however is above the predicted AGN optically thick recombination value of 3.1. This discrepancy is attributed to unaccounted dust extinction at the redshift of our sources diminishing the shorter wavelength H$\beta$ emission. This result provides evidence that it is commonplace for dust to be present in the NLR of Type 2 AGN.

We compare our \oiii{} emission line luminosity measurements to those derived in \citet{reyes2008} of the same sample of spectra to check for consistency. We find a scatter and systematic offset between the measurement methods. Our measurements using the \qsfit{} Type 2 AGN routine presented in this paper are on average 1.16 times larger than those using \citet{reyes2008} semi-parametric method and 1.22 times larger than their Gaussian method. This discrepancy is largely owed to the differing models used for measurements, as we employ a blue wing component when fitting \oiii{} in our method. We show that modifying our routine to use a single Gaussian component for \oiii{} reduces the observed offset to 1.03 and 1.07 for the semi-parametric and Gaussian methods, respectively. The remaining smaller offset is likely explained by the choice of continuum estimation. We employ a 5$\ $Gyr Elliptical host-galaxy template summed with a power-law continuum, whereas a linear fit around the [O{\sc iii}] complex is adopted in \citet{reyes2008} which is free from absorption features in the region.

In our sample we observe that spectra with higher S/N exhibit a higher (statistically worse) $\chi^{2}$ fit statistic than lower S/N spectra. This counter-intuitive trend is attributed to the fact that increased S/N reveals complexities in the emission line profiles that cannot be observed above the noise in lower S/N sources, rendering our modelling of the most prominent emission lines (ie. H$\beta$, \oiii{}, [O{\sc iii}]$\lambda$4959 and H$\alpha$) insufficient. We test a number of emission line profile combinations to better accommodate the modelling of high S/N emission lines and find that using two Gaussians provides an effective solution. Our high S/N ratio spectra recipe is made available as a specialised \qsfit{} recipe for these cases.

We present a general \qsfit{} routine for analysing Type 2 AGN spectra with double peaked narrow emission lines, motivated by the interesting physics in these sources and the frequency at which they are present as outliers in our self-consistency tests. An intersection of 45 spectra is found between \citet{liu2010} sample of double peaked emission line Type 2 AGN and our sample of luminous Type 2 AGN spectra, which we analyse using our specialised double peaked recipe. A distribution of separations between red and blue peaks of double peaked emission lines is observed with a median velocity separation of 390$\pm$60$\ $kms$^{-1}$. The luminosity ratios of blue and red components (L$_{blue}$/L$_{red}$) of double peaked [O {\sc iii}]$\lambda$4959, \oiii{} and H$\alpha$ emission lines are found to be consistent with unity across the sample, showing that it is not favoured for the red or blue peak to be more luminous. Through plotting the red and blue components of our double peaked sample on the BPT emission line diagnostic diagram we show that all lines are consistent with being produced by an AGN ionizing continuum and consistent with the positions of single peaked sources. No evidence is found for the red or blue peaks to have systematically different ionization parameter or metallicity and no sources change classification on the diagram between the two sets of lines, pointing to a common ionization source for each double peaked Type 2 AGN. 

We perform visual inspection for merger evidence in the host-galaxy morphology of Pan-STARRS1 imaging of our double peaked sample and \oiii{} luminosity and redshift matched samples of single peaked AGN from our parent sample. To the resolution available (64$\ $kpc at the median redshift of our sample, $z$=0.28) we find double peaked Type 2 AGN reside in merging systems at a comparable frequency to single peaked AGN across a number of criteria. This suggests the double peaked AGN phenomenon is likely to have a kinematically disturbed NLR origin rather than a dual AGN origin which is expected in late stage mergers in the majority of cases, however this does not eliminate dual AGN as the origin of some of the sources in our sample\footnote{ \url{https://github.com/MattSelwood/Type2AGN}}.

\section*{Acknowledgements}
We thank the anonymous referee for their careful reading of our manuscript and their insightful comments which have served to improve the quality of this work.

This work is supported by the UKRI AIMLAC CDT, funded by grant EP/S023992/1.

Funding for the SDSS and SDSS-II has been provided by the Alfred P. Sloan Foundation, the Participating Institutions, the National Science Foundation, the U.S. Department of Energy, the National Aeronautics and Space Administration, the Japanese Monbukagakusho, the Max Planck Society, and the Higher Education Funding Council for England. The SDSS Web Site is \url{http://www.sdss.org/}.

The SDSS is managed by the Astrophysical Research Consortium for the Participating Institutions. The Participating Institutions are the American Museum of Natural History, Astrophysical Institute Potsdam, University of Basel, University of Cambridge, Case Western Reserve University, University of Chicago, Drexel University, Fermilab, the Institute for Advanced Study, the Japan Participation Group, Johns Hopkins University, the Joint Institute for Nuclear Astrophysics, the Kavli Institute for Particle Astrophysics and Cosmology, the Korean Scientist Group, the Chinese Academy of Sciences (LAMOST), Los Alamos National Laboratory, the Max-Planck-Institute for Astronomy (MPIA), the Max-Planck-Institute for Astrophysics (MPA), New Mexico State University, Ohio State University, University of Pittsburgh, University of Portsmouth, Princeton University, the United States Naval Observatory, and the University of Washington.

The Pan-STARRS1 Surveys (PS1) and the PS1 public science archive have been made possible through contributions by the Institute for Astronomy, the University of Hawaii, the Pan-STARRS Project Office, the Max-Planck Society and its participating institutes, the Max Planck Institute for Astronomy, Heidelberg and the Max Planck Institute for Extraterrestrial Physics, Garching, The Johns Hopkins University, Durham University, the University of Edinburgh, the Queen's University Belfast, the Harvard-Smithsonian Center for Astrophysics, the Las Cumbres Observatory Global Telescope Network Incorporated, the National Central University of Taiwan, the Space Telescope Science Institute, the National Aeronautics and Space Administration under Grant No. NNX08AR22G issued through the Planetary Science Division of the NASA Science Mission Directorate, the National Science Foundation Grant No. AST-1238877, the University of Maryland, Eotvos Lorand University (ELTE), the Los Alamos National Laboratory, and the Gordon and Betty Moore Foundation.

\section*{Data Availability}
The data underlying this article are available in CDS, at https://dx.doi.org/[doi]



\bibliographystyle{mnras}
\bibliography{references} 

\begin{thebibliography}{}
\makeatletter
\relax
\def\mn@urlcharsother{\let\do\@makeother \do\$\do\&\do\#\do\^\do\_\do\%\do\~}
\def\mn@doi{\begingroup\mn@urlcharsother \@ifnextchar [ {\mn@doi@}
  {\mn@doi@[]}}
\def\mn@doi@[#1]#2{\def\@tempa{#1}\ifx\@tempa\@empty \href
  {http://dx.doi.org/#2} {doi:#2}\else \href {http://dx.doi.org/#2} {#1}\fi
  \endgroup}
\def\mn@eprint#1#2{\mn@eprint@#1:#2::\@nil}
\def\mn@eprint@arXiv#1{\href {http://arxiv.org/abs/#1} {{\tt arXiv:#1}}}
\def\mn@eprint@dblp#1{\href {http://dblp.uni-trier.de/rec/bibtex/#1.xml}
  {dblp:#1}}
\def\mn@eprint@#1:#2:#3:#4\@nil{\def\@tempa {#1}\def\@tempb {#2}\def\@tempc
  {#3}\ifx \@tempc \@empty \let \@tempc \@tempb \let \@tempb \@tempa \fi \ifx
  \@tempb \@empty \def\@tempb {arXiv}\fi \@ifundefined
  {mn@eprint@\@tempb}{\@tempb:\@tempc}{\expandafter \expandafter \csname
  mn@eprint@\@tempb\endcsname \expandafter{\@tempc}}}

\bibitem[\protect\citeauthoryear{{Adelman-McCarthy} et~al.,}{{Adelman-McCarthy}
  et~al.}{2008}]{adelmanmccarthy2008}
{Adelman-McCarthy} J.~K.,  et~al., 2008, \mn@doi [\apjs] {10.1086/524984},
  \href {https://ui.adsabs.harvard.edu/abs/2008ApJS..175..297A} {175, 297}

\bibitem[\protect\citeauthoryear{{Agostino} et~al.,}{{Agostino}
  et~al.}{2021}]{agostino2021}
{Agostino} C.~J.,  et~al., 2021, \mn@doi [\apj] {10.3847/1538-4357/ac1e8d},
  \href {https://ui.adsabs.harvard.edu/abs/2021ApJ...922..156A} {922, 156}

\bibitem[\protect\citeauthoryear{{Alexander} \& {Hickox}}{{Alexander} \&
  {Hickox}}{2012}]{alexander&hickox2012}
{Alexander} D.~M.,  {Hickox} R.~C.,  2012, \mn@doi [\nar]
  {10.1016/j.newar.2011.11.003}, \href
  {https://ui.adsabs.harvard.edu/abs/2012NewAR..56...93A} {56, 93}

\bibitem[\protect\citeauthoryear{{Antonucci}}{{Antonucci}}{1993}]{antonucci1993}
{Antonucci} R.,  1993, \mn@doi [\araa] {10.1146/annurev.aa.31.090193.002353},
  \href {https://ui.adsabs.harvard.edu/abs/1993ARA&A..31..473A} {31, 473}

\bibitem[\protect\citeauthoryear{{Baldwin}, {Phillips}  \&
  {Terlevich}}{{Baldwin} et~al.}{1981}]{bpt1981}
{Baldwin} J.~A.,  {Phillips} M.~M.,   {Terlevich} R.,  1981, \mn@doi [\pasp]
  {10.1086/130766}, \href
  {https://ui.adsabs.harvard.edu/abs/1981PASP...93....5B} {93, 5}

\bibitem[\protect\citeauthoryear{Berton}{Berton}{2019}]{berton2019}
Berton M.,  2019, \mn@doi [Proceedings of the International Astronomical Union]
  {10.1017/S1743921320002653}, 15, 94–94

\bibitem[\protect\citeauthoryear{{Bezanson}, {Karpinski}, {Shah}  \&
  {Edelman}}{{Bezanson} et~al.}{2012}]{bazanson2012}
{Bezanson} J.,  {Karpinski} S.,  {Shah} V.~B.,   {Edelman} A.,  2012, arXiv
  e-prints, \href {https://ui.adsabs.harvard.edu/abs/2012arXiv1209.5145B} {p.
  arXiv:1209.5145}

\bibitem[\protect\citeauthoryear{{Blanton} et~al.,}{{Blanton}
  et~al.}{2017}]{blanton2017}
{Blanton} M.~R.,  et~al., 2017, \mn@doi [\aj] {10.3847/1538-3881/aa7567}, \href
  {https://ui.adsabs.harvard.edu/abs/2017AJ....154...28B} {154, 28}

\bibitem[\protect\citeauthoryear{{Bornancini} \& {Garc{\'\i}a
  Lambas}}{{Bornancini} \& {Garc{\'\i}a
  Lambas}}{2020}]{bornancini&garcialambas2020}
{Bornancini} C.,  {Garc{\'\i}a Lambas} D.,  2020, \mn@doi [\mnras]
  {10.1093/mnras/staa723}, \href
  {https://ui.adsabs.harvard.edu/abs/2020MNRAS.494.1189B} {494, 1189}

\bibitem[\protect\citeauthoryear{{Bruzual} \& {Charlot}}{{Bruzual} \&
  {Charlot}}{2003}]{bruzal&charlot2003}
{Bruzual} G.,  {Charlot} S.,  2003, \mn@doi [\mnras]
  {10.1046/j.1365-8711.2003.06897.x}, \href
  {https://ui.adsabs.harvard.edu/abs/2003MNRAS.344.1000B} {344, 1000}

\bibitem[\protect\citeauthoryear{{Calderone}, {Nicastro}, {Ghisellini},
  {Dotti}, {Sbarrato}, {Shankar}  \& {Colpi}}{{Calderone}
  et~al.}{2017}]{calderone2017}
{Calderone} G.,  {Nicastro} L.,  {Ghisellini} G.,  {Dotti} M.,  {Sbarrato} T.,
  {Shankar} F.,   {Colpi} M.,  2017, \mn@doi [\mnras] {10.1093/mnras/stx2239},
  \href {https://ui.adsabs.harvard.edu/abs/2017MNRAS.472.4051C} {472, 4051}

\bibitem[\protect\citeauthoryear{{Chambers} et~al.,}{{Chambers}
  et~al.}{2016}]{chambers2016}
{Chambers} K.~C.,  et~al., 2016, arXiv e-prints, \href
  {https://ui.adsabs.harvard.edu/abs/2016arXiv161205560C} {p. arXiv:1612.05560}

\bibitem[\protect\citeauthoryear{{Cirasuolo} et~al.,}{{Cirasuolo}
  et~al.}{2020}]{cirasuolo2020}
{Cirasuolo} M.,  et~al., 2020, \mn@doi [The Messenger]
  {10.18727/0722-6691/5195}, \href
  {https://ui.adsabs.harvard.edu/abs/2020Msngr.180...10C} {180, 10}

\bibitem[\protect\citeauthoryear{{Comerford}, {Schluns}, {Greene}  \&
  {Cool}}{{Comerford} et~al.}{2013}]{comerford2013}
{Comerford} J.~M.,  {Schluns} K.,  {Greene} J.~E.,   {Cool} R.~J.,  2013,
  \mn@doi [\apj] {10.1088/0004-637X/777/1/64}, \href
  {https://ui.adsabs.harvard.edu/abs/2013ApJ...777...64C} {777, 64}

\bibitem[\protect\citeauthoryear{{DESI Collaboration} et~al.,}{{DESI
  Collaboration} et~al.}{2016}]{desi2016}
{DESI Collaboration} et~al., 2016, arXiv e-prints, \href
  {https://ui.adsabs.harvard.edu/abs/2016arXiv161100036D} {p. arXiv:1611.00036}

\bibitem[\protect\citeauthoryear{{Dey} et~al.,}{{Dey} et~al.}{2019}]{dey2019}
{Dey} A.,  et~al., 2019, \mn@doi [\aj] {10.3847/1538-3881/ab089d}, \href
  {https://ui.adsabs.harvard.edu/abs/2019AJ....157..168D} {157, 168}

\bibitem[\protect\citeauthoryear{{Dias dos Santos}, {Rodrguez-Ardila}  \&
  {Marinello}}{{Dias dos Santos} et~al.}{2022}]{diasdossantos2022}
{Dias dos Santos} D.,  {Rodrguez-Ardila} A.,   {Marinello} M.,  2022, \mn@doi
  [Astronomische Nachrichten] {10.1002/asna.20210098}, \href
  {https://ui.adsabs.harvard.edu/abs/2022AN....34310098D} {343, e210098}

\bibitem[\protect\citeauthoryear{{Dom{\'\i}nguez} et~al.,}{{Dom{\'\i}nguez}
  et~al.}{2013}]{dominguez2013}
{Dom{\'\i}nguez} A.,  et~al., 2013, \mn@doi [\apj]
  {10.1088/0004-637X/763/2/145}, \href
  {https://ui.adsabs.harvard.edu/abs/2013ApJ...763..145D} {763, 145}

\bibitem[\protect\citeauthoryear{{Francis}, {Hewett}, {Foltz}, {Chaffee},
  {Weymann}  \& {Morris}}{{Francis} et~al.}{1991}]{francis1991}
{Francis} P.~J.,  {Hewett} P.~C.,  {Foltz} C.~B.,  {Chaffee} F.~H.,  {Weymann}
  R.~J.,   {Morris} S.~L.,  1991, \mn@doi [\apj] {10.1086/170066}, \href
  {https://ui.adsabs.harvard.edu/abs/1991ApJ...373..465F} {373, 465}

\bibitem[\protect\citeauthoryear{{Gaskell}}{{Gaskell}}{2017}]{gaskell2017}
{Gaskell} C.~M.,  2017, \mn@doi [\mnras] {10.1093/mnras/stx094}, \href
  {https://ui.adsabs.harvard.edu/abs/2017MNRAS.467..226G} {467, 226}

\bibitem[\protect\citeauthoryear{Gaskell \& Ferland}{Gaskell \&
  Ferland}{1984}]{gaskell&ferland1984}
Gaskell C.~M.,  Ferland G.~J.,  1984, \mn@doi [Publications of the Astronomical
  Society of the Pacific] {10.1086/131352}, 96, 393

\bibitem[\protect\citeauthoryear{{Halpern} \& {Steiner}}{{Halpern} \&
  {Steiner}}{1983}]{halpern&steiner1983}
{Halpern} J.~P.,  {Steiner} J.~E.,  1983, \mn@doi [\apjl] {10.1086/184051},
  \href {https://ui.adsabs.harvard.edu/abs/1983ApJ...269L..37H} {269, L37}

\bibitem[\protect\citeauthoryear{{Hickox} \& {Alexander}}{{Hickox} \&
  {Alexander}}{2018}]{hickox&alexander2018}
{Hickox} R.~C.,  {Alexander} D.~M.,  2018, \mn@doi [\araa]
  {10.1146/annurev-astro-081817-051803}, \href
  {https://ui.adsabs.harvard.edu/abs/2018ARA&A..56..625H} {56, 625}

\bibitem[\protect\citeauthoryear{{Ilbert} et~al.,}{{Ilbert}
  et~al.}{2009}]{ilbert2009}
{Ilbert} O.,  et~al., 2009, \mn@doi [\apj] {10.1088/0004-637X/690/2/1236},
  \href {https://ui.adsabs.harvard.edu/abs/2009ApJ...690.1236I} {690, 1236}

\bibitem[\protect\citeauthoryear{{Jin}, {Ward}  \& {Done}}{{Jin}
  et~al.}{2012}]{jin2012}
{Jin} C.,  {Ward} M.,   {Done} C.,  2012, \mn@doi [\mnras]
  {10.1111/j.1365-2966.2012.20847.x}, \href
  {https://ui.adsabs.harvard.edu/abs/2012MNRAS.422.3268J} {422, 3268}

\bibitem[\protect\citeauthoryear{{Kauffmann} et~al.,}{{Kauffmann}
  et~al.}{2003}]{kauffmann2003}
{Kauffmann} G.,  et~al., 2003, \mn@doi [\mnras]
  {10.1111/j.1365-2966.2003.07154.x}, \href
  {https://ui.adsabs.harvard.edu/abs/2003MNRAS.346.1055K} {346, 1055}

\bibitem[\protect\citeauthoryear{{Kewley}, {Dopita}, {Sutherland}, {Heisler}
  \& {Trevena}}{{Kewley} et~al.}{2001}]{kewley2001}
{Kewley} L.~J.,  {Dopita} M.~A.,  {Sutherland} R.~S.,  {Heisler} C.~A.,
  {Trevena} J.,  2001, \mn@doi [\apj] {10.1086/321545}, \href
  {https://ui.adsabs.harvard.edu/abs/2001ApJ...556..121K} {556, 121}

\bibitem[\protect\citeauthoryear{{Kewley}, {Groves}, {Kauffmann}  \&
  {Heckman}}{{Kewley} et~al.}{2006}]{kewley2006}
{Kewley} L.~J.,  {Groves} B.,  {Kauffmann} G.,   {Heckman} T.,  2006, \mn@doi
  [\mnras] {10.1111/j.1365-2966.2006.10859.x}, \href
  {https://ui.adsabs.harvard.edu/abs/2006MNRAS.372..961K} {372, 961}

\bibitem[\protect\citeauthoryear{{Kochanek} et~al.,}{{Kochanek}
  et~al.}{2012}]{kochanek2012}
{Kochanek} C.~S.,  et~al., 2012, \mn@doi [\apjs] {10.1088/0067-0049/200/1/8},
  \href {https://ui.adsabs.harvard.edu/abs/2012ApJS..200....8K} {200, 8}

\bibitem[\protect\citeauthoryear{{Kollatschny} \& {Zetzl}}{{Kollatschny} \&
  {Zetzl}}{2013}]{kollatshny2013}
{Kollatschny} W.,  {Zetzl} M.,  2013, \mn@doi [\aap]
  {10.1051/0004-6361/201219411}, \href
  {https://ui.adsabs.harvard.edu/abs/2013A&A...549A.100K} {549, A100}

\bibitem[\protect\citeauthoryear{Kuraszkiewicz, Green, Crenshaw, Dunn, Forster,
  Vestergaard  \& Aldcroft}{Kuraszkiewicz et~al.}{2004}]{kuraszkiewicz2004}
Kuraszkiewicz J.~K.,  Green P.~J.,  Crenshaw D.~M.,  Dunn J.,  Forster K.,
  Vestergaard M.,   Aldcroft T.~L.,  2004, \mn@doi [The Astrophysical Journal
  Supplement Series] {10.1086/379809}, 150, 165

\bibitem[\protect\citeauthoryear{{Lacy}, {Ridgway}, {Sajina}, {Petric},
  {Gates}, {Urrutia}  \& {Storrie-Lombardi}}{{Lacy} et~al.}{2015}]{lacy2015}
{Lacy} M.,  {Ridgway} S.~E.,  {Sajina} A.,  {Petric} A.~O.,  {Gates} E.~L.,
  {Urrutia} T.,   {Storrie-Lombardi} L.~J.,  2015, \mn@doi [\apj]
  {10.1088/0004-637X/802/2/102}, \href
  {https://ui.adsabs.harvard.edu/abs/2015ApJ...802..102L} {802, 102}

\bibitem[\protect\citeauthoryear{{Lamastra}, {Bianchi}, {Matt}, {Perola},
  {Barcons}  \& {Carrera}}{{Lamastra} et~al.}{2009}]{lamastra2009}
{Lamastra} A.,  {Bianchi} S.,  {Matt} G.,  {Perola} G.~C.,  {Barcons} X.,
  {Carrera} F.~J.,  2009, \mn@doi [\aap] {10.1051/0004-6361/200912023}, \href
  {https://ui.adsabs.harvard.edu/abs/2009A&A...504...73L} {504, 73}

\bibitem[\protect\citeauthoryear{{Laureijs} et~al.,}{{Laureijs}
  et~al.}{2011}]{laureijs2011}
{Laureijs} R.,  et~al., 2011, arXiv e-prints, \href
  {https://ui.adsabs.harvard.edu/abs/2011arXiv1110.3193L} {p. arXiv:1110.3193}

\bibitem[\protect\citeauthoryear{{Lehmann} et~al.,}{{Lehmann}
  et~al.}{2001}]{lehmann2001}
{Lehmann} I.,  et~al., 2001, \mn@doi [\aap] {10.1051/0004-6361:20010419}, \href
  {https://ui.adsabs.harvard.edu/abs/2001A&A...371..833L} {371, 833}

\bibitem[\protect\citeauthoryear{{Liu}, {Shen}, {Strauss}  \& {Greene}}{{Liu}
  et~al.}{2010}]{liu2010}
{Liu} X.,  {Shen} Y.,  {Strauss} M.~A.,   {Greene} J.~E.,  2010, \mn@doi [\apj]
  {10.1088/0004-637X/708/1/427}, \href
  {https://ui.adsabs.harvard.edu/abs/2010ApJ...708..427L} {708, 427}

\bibitem[\protect\citeauthoryear{{Lu}, {Zhao}, {Bai}  \& {Fan}}{{Lu}
  et~al.}{2019}]{lu2019}
{Lu} K.-X.,  {Zhao} Y.,  {Bai} J.-M.,   {Fan} X.-L.,  2019, \mn@doi [\mnras]
  {10.1093/mnras/sty3229}, \href
  {https://ui.adsabs.harvard.edu/abs/2019MNRAS.483.1722L} {483, 1722}

\bibitem[\protect\citeauthoryear{{Markwardt}}{{Markwardt}}{2009}]{markwardt2009}
{Markwardt} C.~B.,  2009, in {Bohlender} D.~A.,  {Durand} D.,   {Dowler} P.,
  eds,  Astronomical Society of the Pacific Conference Series Vol. 411,
  Astronomical Data Analysis Software and Systems XVIII. p.~251 (\mn@eprint
  {arXiv} {0902.2850})

\bibitem[\protect\citeauthoryear{{Marziani}, {Sulentic}, {Plauchu-Frayn}  \&
  {del Olmo}}{{Marziani} et~al.}{2013}]{marziani2013}
{Marziani} P.,  {Sulentic} J.~W.,  {Plauchu-Frayn} I.,   {del Olmo} A.,  2013,
  \mn@doi [\aap] {10.1051/0004-6361/201321374}, \href
  {https://ui.adsabs.harvard.edu/abs/2013A&A...555A..89M} {555, A89}

\bibitem[\protect\citeauthoryear{{Merloni} et~al.,}{{Merloni}
  et~al.}{2019}]{merloni2019}
{Merloni} A.,  et~al., 2019, \mn@doi [The Messenger] {10.18727/0722-6691/5125},
  \href {https://ui.adsabs.harvard.edu/abs/2019Msngr.175...42M} {175, 42}

\bibitem[\protect\citeauthoryear{{Naddaf} \& {Czerny}}{{Naddaf} \&
  {Czerny}}{2021}]{naddaf2021}
{Naddaf} M.-H.,  {Czerny} B.,  2021, arXiv e-prints, \href
  {https://ui.adsabs.harvard.edu/abs/2021arXiv211114963N} {p. arXiv:2111.14963}

\bibitem[\protect\citeauthoryear{{Netzer}}{{Netzer}}{2015}]{netzer2015}
{Netzer} H.,  2015, \mn@doi [\araa] {10.1146/annurev-astro-082214-122302},
  \href {https://ui.adsabs.harvard.edu/abs/2015ARA&A..53..365N} {53, 365}

\bibitem[\protect\citeauthoryear{{O'Donnell}}{{O'Donnell}}{1994}]{odonnell1994}
{O'Donnell} J.~E.,  1994, \mn@doi [\apj] {10.1086/173713}, \href
  {https://ui.adsabs.harvard.edu/abs/1994ApJ...422..158O} {422, 158}

\bibitem[\protect\citeauthoryear{{Osterbrock} \& {Ferland}}{{Osterbrock} \&
  {Ferland}}{2006}]{osterbrock1989}
{Osterbrock} D.~E.,  {Ferland} G.~J.,  2006, {Astrophysics of gaseous nebulae
  and active galactic nuclei}

\bibitem[\protect\citeauthoryear{{Polletta} et~al.,}{{Polletta}
  et~al.}{2007}]{polletta2007}
{Polletta} M.,  et~al., 2007, \mn@doi [\apj] {10.1086/518113}, \href
  {https://ui.adsabs.harvard.edu/abs/2007ApJ...663...81P} {663, 81}

\bibitem[\protect\citeauthoryear{{Predehl} \& {Schmitt}}{{Predehl} \&
  {Schmitt}}{1995}]{predehl&schmitt1995}
{Predehl} P.,  {Schmitt} J.~H.~M.~M.,  1995, \aap, \href
  {https://ui.adsabs.harvard.edu/abs/1995A&A...293..889P} {500, 459}

\bibitem[\protect\citeauthoryear{{Puchnarewicz} et~al.,}{{Puchnarewicz}
  et~al.}{1997}]{purchnarewicz1997}
{Puchnarewicz} E.~M.,  et~al., 1997, \mn@doi [\mnras]
  {10.1093/mnras/291.1.177}, \href
  {https://ui.adsabs.harvard.edu/abs/1997MNRAS.291..177P} {291, 177}

\bibitem[\protect\citeauthoryear{{Ramos Almeida} \& {Ricci}}{{Ramos Almeida} \&
  {Ricci}}{2017}]{ramosalmeida2017}
{Ramos Almeida} C.,  {Ricci} C.,  2017, \mn@doi [Nature Astronomy]
  {10.1038/s41550-017-0232-z}, \href
  {https://ui.adsabs.harvard.edu/abs/2017NatAs...1..679R} {1, 679}

\bibitem[\protect\citeauthoryear{{Reyes} et~al.,}{{Reyes}
  et~al.}{2008}]{reyes2008}
{Reyes} R.,  et~al., 2008, \mn@doi [\aj] {10.1088/0004-6256/136/6/2373}, \href
  {https://ui.adsabs.harvard.edu/abs/2008AJ....136.2373R} {136, 2373}

\bibitem[\protect\citeauthoryear{{Sarzi} et~al.,}{{Sarzi}
  et~al.}{2006}]{sarzi2006}
{Sarzi} M.,  et~al., 2006, \mn@doi [\mnras] {10.1111/j.1365-2966.2005.09839.x},
  \href {https://ui.adsabs.harvard.edu/abs/2006MNRAS.366.1151S} {366, 1151}

\bibitem[\protect\citeauthoryear{{Schlegel}, {Finkbeiner}  \&
  {Davis}}{{Schlegel} et~al.}{1998}]{schlegel1998}
{Schlegel} D.~J.,  {Finkbeiner} D.~P.,   {Davis} M.,  1998, \mn@doi [\apj]
  {10.1086/305772}, \href
  {https://ui.adsabs.harvard.edu/abs/1998ApJ...500..525S} {500, 525}

\bibitem[\protect\citeauthoryear{{Schmidt}}{{Schmidt}}{1968}]{schmidt1968}
{Schmidt} M.,  1968, \mn@doi [\apj] {10.1086/149446}, \href
  {https://ui.adsabs.harvard.edu/abs/1968ApJ...151..393S} {151, 393}

\bibitem[\protect\citeauthoryear{{Schnorr-M{\"u}ller}
  et~al.,}{{Schnorr-M{\"u}ller} et~al.}{2016}]{schnorr-muller2016}
{Schnorr-M{\"u}ller} A.,  et~al., 2016, \mn@doi [\mnras]
  {10.1093/mnras/stw1865}, \href
  {https://ui.adsabs.harvard.edu/abs/2016MNRAS.462.3570S} {462, 3570}

\bibitem[\protect\citeauthoryear{{Sexton}, {Matzko}, {Darden}, {Canalizo}  \&
  {Gorjian}}{{Sexton} et~al.}{2021}]{sexton2021}
{Sexton} R.~O.,  {Matzko} W.,  {Darden} N.,  {Canalizo} G.,   {Gorjian} V.,
  2021, \mn@doi [\mnras] {10.1093/mnras/staa3278}, \href
  {https://ui.adsabs.harvard.edu/abs/2021MNRAS.500.2871S} {500, 2871}

\bibitem[\protect\citeauthoryear{{Shen}, {Liu}, {Greene}  \& {Strauss}}{{Shen}
  et~al.}{2011}]{shen2011}
{Shen} Y.,  {Liu} X.,  {Greene} J.~E.,   {Strauss} M.~A.,  2011, \mn@doi [\apj]
  {10.1088/0004-637X/735/1/48}, \href
  {https://ui.adsabs.harvard.edu/abs/2011ApJ...735...48S} {735, 48}

\bibitem[\protect\citeauthoryear{{Silva}, {Granato}, {Bressan}  \&
  {Danese}}{{Silva} et~al.}{1998}]{silva1998}
{Silva} L.,  {Granato} G.~L.,  {Bressan} A.,   {Danese} L.,  1998, \mn@doi
  [\apj] {10.1086/306476}, \href
  {https://ui.adsabs.harvard.edu/abs/1998ApJ...509..103S} {509, 103}

\bibitem[\protect\citeauthoryear{{Smith}, {Shields}, {Bonning}, {McMullen},
  {Rosario}  \& {Salviander}}{{Smith} et~al.}{2010}]{smith2010}
{Smith} K.~L.,  {Shields} G.~A.,  {Bonning} E.~W.,  {McMullen} C.~C.,
  {Rosario} D.~J.,   {Salviander} S.,  2010, \mn@doi [\apj]
  {10.1088/0004-637X/716/1/866}, \href
  {https://ui.adsabs.harvard.edu/abs/2010ApJ...716..866S} {716, 866}

\bibitem[\protect\citeauthoryear{{Sulentic}, {Marziani}, {Zamanov}, {Bachev},
  {Calvani}  \& {Dultzin-Hacyan}}{{Sulentic} et~al.}{2002}]{sulentic2002}
{Sulentic} J.~W.,  {Marziani} P.,  {Zamanov} R.,  {Bachev} R.,  {Calvani} M.,
  {Dultzin-Hacyan} D.,  2002, \mn@doi [\apjl] {10.1086/339594}, \href
  {https://ui.adsabs.harvard.edu/abs/2002ApJ...566L..71S} {566, L71}

\bibitem[\protect\citeauthoryear{{Urry} \& {Padovani}}{{Urry} \&
  {Padovani}}{1995}]{urry&padovani1995}
{Urry} C.~M.,  {Padovani} P.,  1995, \mn@doi [\pasp] {10.1086/133630}, \href
  {https://ui.adsabs.harvard.edu/abs/1995PASP..107..803U} {107, 803}

\bibitem[\protect\citeauthoryear{{Vanden Berk} et~al.,}{{Vanden Berk}
  et~al.}{2001}]{vandenberk2001}
{Vanden Berk} D.~E.,  et~al., 2001, \mn@doi [\aj] {10.1086/321167}, \href
  {https://ui.adsabs.harvard.edu/abs/2001AJ....122..549V} {122, 549}

\bibitem[\protect\citeauthoryear{{Veilleux}, {Cecil}  \&
  {Bland-Hawthorn}}{{Veilleux} et~al.}{2005}]{veilleux2005}
{Veilleux} S.,  {Cecil} G.,   {Bland-Hawthorn} J.,  2005, \mn@doi [\araa]
  {10.1146/annurev.astro.43.072103.150610}, \href
  {https://ui.adsabs.harvard.edu/abs/2005ARA&A..43..769V} {43, 769}

\bibitem[\protect\citeauthoryear{{V{\'e}ron-Cetty}, {V{\'e}ron}  \&
  {Gon{\c{c}}alves}}{{V{\'e}ron-Cetty} et~al.}{2001}]{veroncetty2001}
{V{\'e}ron-Cetty} M.~P.,  {V{\'e}ron} P.,   {Gon{\c{c}}alves} A.~C.,  2001,
  \mn@doi [\aap] {10.1051/0004-6361:20010489}, \href
  {https://ui.adsabs.harvard.edu/abs/2001A&A...372..730V} {372, 730}

\bibitem[\protect\citeauthoryear{{Villarroel} \& {Korn}}{{Villarroel} \&
  {Korn}}{2014}]{villarroel&korn2014}
{Villarroel} B.,  {Korn} A.~J.,  2014, \mn@doi [Nature Physics]
  {10.1038/nphys2951}, \href
  {https://ui.adsabs.harvard.edu/abs/2014NatPh..10..417V} {10, 417}

\bibitem[\protect\citeauthoryear{{Wang}, {Chen}, {Hu}, {Mao}, {Zhang}  \&
  {Bian}}{{Wang} et~al.}{2009}]{wang2009}
{Wang} J.-M.,  {Chen} Y.-M.,  {Hu} C.,  {Mao} W.-M.,  {Zhang} S.,   {Bian}
  W.-H.,  2009, \mn@doi [\apjl] {10.1088/0004-637X/705/1/L76}, \href
  {https://ui.adsabs.harvard.edu/abs/2009ApJ...705L..76W} {705, L76}

\bibitem[\protect\citeauthoryear{{Wang} et~al.,}{{Wang}
  et~al.}{2019}]{wang2019}
{Wang} M.~X.,  et~al., 2019, \mn@doi [\mnras] {10.1093/mnras/sty2818}, \href
  {https://ui.adsabs.harvard.edu/abs/2019MNRAS.482.1889W} {482, 1889}

\bibitem[\protect\citeauthoryear{{Zakamska} \& {Greene}}{{Zakamska} \&
  {Greene}}{2014}]{zakamska2014}
{Zakamska} N.~L.,  {Greene} J.~E.,  2014, \mn@doi [\mnras]
  {10.1093/mnras/stu842}, \href
  {https://ui.adsabs.harvard.edu/abs/2014MNRAS.442..784Z} {442, 784}

\bibitem[\protect\citeauthoryear{{Zakamska} et~al.,}{{Zakamska}
  et~al.}{2003}]{zakamska2003}
{Zakamska} N.~L.,  et~al., 2003, \mn@doi [\aj] {10.1086/378610}, \href
  {https://ui.adsabs.harvard.edu/abs/2003AJ....126.2125Z} {126, 2125}

\bibitem[\protect\citeauthoryear{{Zou}, {Yang}, {Brandt}  \& {Xue}}{{Zou}
  et~al.}{2019}]{zou2019}
{Zou} F.,  {Yang} G.,  {Brandt} W.~N.,   {Xue} Y.,  2019, \mn@doi [\apj]
  {10.3847/1538-4357/ab1eb1}, \href
  {https://ui.adsabs.harvard.edu/abs/2019ApJ...878...11Z} {878, 11}

\bibitem[\protect\citeauthoryear{{de Jong} et~al.,}{{de Jong}
  et~al.}{2019}]{dejong2019}
{de Jong} R.~S.,  et~al., 2019, \mn@doi [The Messenger]
  {10.18727/0722-6691/5117}, \href
  {https://ui.adsabs.harvard.edu/abs/2019Msngr.175....3D} {175, 3}

\makeatother
\end{thebibliography}



\appendix
\section{Description of Catalogues}

\subsection{Luminous Type 2 AGN Catalogue}\label{app:cat_descr}
We provide a catalogue of our \qsfit{} measurements of the spectral properties of our sample of 813 luminous Type 2 AGN. The catalogue is deposited as a {\sc FITS} table with 200 columns and 813 rows, available on CDS.

In this section we will list the name and meaning of each column in the \qsfit{} catalogue. Throughout the catalogue the value \verb|-999.0| is used as a fill value for when there is no measurement available for a quantity in a given spectrum. Parameters that are patched to another to provide a single free parameter fit are provided in the catalogue with a null value for their uncertainty and inherit the uncertainty on the parameter they are patched to.

\begin{table}
 \caption{List of emission lines considered in our Type 2 AGN \qsfit{} recipe, their rest-frame vacuum wavelength and their corresponding label in the \qsfit{} catalogue provided alongside this work.}
 \label{tab:cat_labels}
 \centering
 \begin{tabular}{lcc}
  \hline
  Line & Wavelength (\AA) & Label \\
  \hline
  Mg {\sc ii} & 2796.35 & \verb|MgII_2798| \\[2pt]
  [Ne {\sc v}] & 3426.50 & \verb|NeV_3426| \\[2pt]
  [{\sc O ii}] & 3728.48 & \verb|OII_3727| \\[2pt]
  [Ne {\sc iii}] & 3870.16 & \verb|NeIII_3869| \\[2pt]
  H$\gamma$ & 4341.68 & \verb|Hg| \\[2pt]
  H$\beta$ & 4862.68 & \verb|Hb| \\[2pt]
  [{\sc O iii}] & 4960.30 & \verb|OIII_4959| \\[2pt]
  [{\sc O iii}]bw & 4960.30 & \verb|OIII_4959_bw| \\[2pt]
  [{\sc O iii}] & 5008.24 & \verb|OIII_5007| \\[2pt]
  [{\sc O iii}]bw & 5008.24 & \verb|OIII_5007_bw| \\[2pt]
  [{\sc O i}] & 6302.05 & \verb|OI_6300| \\[2pt]
  [{\sc O i}] & 6365.54 & \verb|OI_6364| \\[2pt]
  [{\sc N ii}] & 6549.85 & \verb|NII_6549| \\[2pt]
  H$\alpha$ & 6564.61 & \verb|Ha| \\[2pt]
  [{\sc N ii}] & 6585.28 & \verb|NII_6583| \\[2pt]
  [{\sc S ii}] & 6718.29 & \verb|SII_6716| \\[2pt]
  [{\sc S ii}] & 6732.67 & \verb|SII_6731| \\[2pt]
  Unknown 1 & - & \verb|unk1| \\[2pt]
  Unknown 2 & - & \verb|unk2| \\[2pt]
  \hline
 \end{tabular}
\end{table}

The list of columns provided in our catalogue are as follows:

\begin{enumerate}
    \item \verb|SOURCE_ID|: name of the SDSS spectrum given in the format `spec-PLATE-MJD-FIBER';
    \item \verb|SDSS_ID|: SDSS object name of the source given by its right ascension and declination co-ordinates in the format `SDSS JHHMMSS.ss+DDMMSS.s';
    \item \verb|RA|: the right ascension (J2000.0) of the source;
    \item \verb|DEC|: the declination (J2000.0) of the source;
    \item \verb|Z_OBS|: the redshift of the source;
    \item \verb|SN|: the median S/N of the spectrum;
    \item \verb|EBV|: the colour excess used to de-redden the spectrum prior to fitting;
    \item \verb|RECIPE|: the name of the \qsfit{} recipe used to analyse the spectrum;
    \item \verb|model_elapsed|: time elapsed to complete fitting of the spectrum, given in units of seconds;
    \item \verb|model_ndata|: the number of data points in the spectrum;
    \item \verb|model_nfree|: the number of free parameters in the fit;
    \item \verb|model_dof|: the number of degrees of freedom in the fit;
    \item \verb|model_fitstat|: the total $\chi^{2}$ fit statistic of the model;
    \item \verb|qso_cont_norm|, \verb|qso_cont_norm_unc|: the $\lambda$L$_{\lambda}$ luminosity of the power-law continuum at the reference wavelength of the component and its associated statistical uncertainty, given in units of 10$^{42}\ $erg~s$^{-1}$;
    \item \verb|qso_cont_x0|: reference wavelength of the power-law continuum fixed at median wavelength value of the spectrum, given in units of \AA;
    \item \verb|qso_cont_alpha|, \verb|qso_cont_alpha_unc|: power-law continuum slope and associated statistical uncertainty;
    \item \verb|galaxy_norm|, \verb|galaxy_norm_unc|: the $\lambda$L$_{\lambda}$ luminosity of the host-galaxy template at 5500~\AA, given in units of 10$^{42}\ $erg~s$^{-1}$;
    \item The remaining columns pertain to emission line properties. Each property holds the same meaning for each of the lines considered in the recipe. The properties associated with each emission line is denoted using the assigned labels presented in Table \ref{tab:cat_labels} which should be inserted in the \verb|[LINE_ID]| part of the column name.
    \begin{enumerate}
        \item \verb|[LINE_ID]_norm|, \verb|[LINE_ID]_norm_unc|: integrated luminosity of the emission line profile and its associated statistical uncertainty, given in units of 10$^{42}\ $erg~s$^{-1}$;
        \item \verb|[LINE_ID]_center|: central rest-frame wavelength of the emission line, given in units of \AA;
        \item \verb|[LINE_ID]_fwhm|, \verb|[LINE_ID]_fwhm_unc|: full width at half maximum of the fitted emission line profile and its associated statistical uncertainty, given in units of km~s$^{-1}$;
        \item \verb|[LINE_ID]_voff|, \verb|[LINE_ID]_voff_unc|: velocity offset of the emission line from the system velocity and its associated statistical uncertainty, given in units of km~s$^{-1}$;
        \item \verb|[LINE_ID]_ew|, \verb|[LINE_ID]_ew_unc|: equivalent width of the emission line and its associated statistical uncertainty, given in units of \AA.
    \end{enumerate}
    
\end{enumerate}

\subsection{Double Peaked Type 2 AGN Catalogue}\label{app:dpcat}

Alongside our catalogue of measurements for our main single peaked luminous Type 2 AGN sample, we publish a catalogue of measurements for our double peaked Type 2 AGN sub-sample. Our double peaked catalogue is depositied as a {\sc FITS} table with 263 columns and 45 rows, available on CDS.

All column names and meanings are identical to those described in the prior section, however with a revised emission line list containing the components of our double peaked Type 2 AGN recipe which is given in Table \ref{tab:dp_cat_labels}.

\begin{table}
 \caption{List of emission lines considered in our double peaked Type 2 AGN \qsfit{} recipe, their rest-frame vacuum wavelength and their corresponding label in the double peaked \qsfit{} catalogue provided alongside this work.}
 \label{tab:dp_cat_labels}
 \centering
 \begin{tabular}{lcc}
  \hline
  Line & Wavelength (\AA) & Label \\
  \hline
  Mg {\sc ii} & 2796.35 & \verb|MgII_2798| \\[2pt]
  [Ne {\sc v}] & 3426.50 & \verb|NeV_3426| \\[2pt]
  [{\sc O ii}]$_{blue}$ & 3728.48 & \verb|OII_3727_2| \\[2pt]
  [{\sc O ii}]$_{red}$ & 3728.48 & \verb|OII_3727| \\[2pt]
  [Ne {\sc iii}] & 3870.16 & \verb|NeIII_3869| \\[2pt]
  H$\gamma$ & 4341.68 & \verb|Hg| \\[2pt]
  H$\beta_{blue}$ & 4862.68 & \verb|Hb_2| \\[2pt]
  H$\beta_{red}$ & 4862.68 & \verb|Hb| \\[2pt]
  [{\sc O iii}]$_{blue}$ & 4960.30 & \verb|OIII_4959_2| \\[2pt]
  [{\sc O iii}]$_{red}$ & 4960.30 & \verb|OIII_4959| \\[2pt]
  [{\sc O iii}]$_{core}$ & 4960.30 & \verb|OIII_4959_core| \\[2pt]
  [{\sc O iii}]$_{blue}$ & 5008.24 & \verb|OIII_5007_2| \\[2pt]
  [{\sc O iii}]$_{red}$ & 5008.24 & \verb|OIII_5007| \\[2pt]
  [{\sc O iii}]$_{core}$ & 5008.24 & \verb|OIII_5007_core| \\[2pt]
  [{\sc O i}] & 6302.05 & \verb|OI_6300| \\[2pt]
  [{\sc O i}] & 6365.54 & \verb|OI_6364| \\[2pt]
  [{\sc N ii}] & 6549.85 & \verb|NII_6549| \\[2pt]
  H$\alpha_{blue}$ & 6564.61 & \verb|Ha_2| \\[2pt]
  H$\alpha_{red}$ & 6564.61 & \verb|Ha| \\[2pt]
  [{\sc N ii}]$_{blue}$ & 6585.28 & \verb|NII_6583_2| \\[2pt]
  [{\sc N ii}]$_{red}$ & 6585.28 & \verb|NII_6583| \\[2pt]
  [{\sc S ii}] & 6718.29 & \verb|SII_6716| \\[2pt]
  [{\sc S ii}] & 6732.67 & \verb|SII_6731| \\[2pt]
  Unknown 1 & - & \verb|unk1| \\[2pt]
  Unknown 2 & - & \verb|unk2| \\[2pt]
  Unknown 3 & - & \verb|unk3| \\[2pt]
  Unknown 4 & - & \verb|unk4| \\[2pt]
  \hline
 \end{tabular}
\end{table}


\bsp	
\label{lastpage}
\end{document}